\documentclass[aps,prb,twocolumn,superscriptaddress]{revtex4-2}
\usepackage[T1]{fontenc}

\usepackage{graphicx,epstopdf,siunitx,braket,xspace,amssymb,dsfont,bm,amsmath,float}
\usepackage{comment}
\usepackage[linktocpage=true,colorlinks=true,urlcolor=blue,linkcolor=red,citecolor = blue]{hyperref}
\usepackage{pgfplots}
\usepackage{amsfonts}
\usepackage{xcolor}

\newcommand{\DU}[1]{\ensuremath{\ket{\downarrow \uparrow}}}

\begin{document}

\title{Fast high-fidelity baseband reset of a latched state for quantum dot qubit readout}

\author{Piotr Marciniec}
\altaffiliation[These authors ]{contributed equally}
\affiliation{Department of Physics, University of Wisconsin-Madison, Madison, WI 53706, USA}

\author{M. A. Wolfe}
\altaffiliation[These authors ]{contributed equally}
\affiliation{Department of Physics, University of Wisconsin-Madison, Madison, WI 53706, USA}

\author{Tyler Kovach}
\affiliation{Department of Physics, University of Wisconsin-Madison, Madison, WI 53706, USA}

\author{J. Reily}
\affiliation{Department of Physics, University of Wisconsin-Madison, Madison, WI 53706, USA}

\author{Sanghyeok Park}
\affiliation{Department of Physics, University of Wisconsin-Madison, Madison, WI 53706, USA}

\author{Jared Benson}
\affiliation{Department of Physics, University of Wisconsin-Madison, Madison, WI 53706, USA}

\author{Mark Friesen}
\affiliation{Department of Physics, University of Wisconsin-Madison, Madison, WI 53706, USA}

\author{Benjamin D. Woods}
\affiliation{Department of Physics, University of Wisconsin-Madison, Madison, WI 53706, USA}

\author{Matthew J. Curry}
\affiliation{Intel Corp., Hillsboro, OR, USA}

\author{Nathaniel C. Bishop}
\affiliation{Intel Corp., Hillsboro, OR, USA}

\author{J. Corrigan}
\affiliation{Intel Corp., Hillsboro, OR, USA}

\author{M. A. Eriksson}
\affiliation{Department of Physics, University of Wisconsin-Madison, Madison, WI 53706, USA}

\begin{abstract}
A common method for reading out the state of a spin qubit is by latching one logical qubit state, either $\ket{1}$ or $\ket{0}$, onto a different, metastable charge state. Such a latched state can provide a superior charge sensing signal for qubit readout, and it can have a lifetime chosen to be long enough that the charge sensed readout can be high fidelity. However, the passive reset out of latched states is inherently long, which is not desirable. In this work, we demonstrate an on-demand, high fidelity ($>99\%$) re-initialization of a quantum dot qubit out of a latched readout state. The method is simple to apply as it involves a single baseband voltage pulse to a specific region in the quantum dot stability diagram where the relaxation time from the latched state to the ground state is over 50 times faster. We describe the mechanism for the reset process as well as the boundaries for the optimal reset region in the qubit gate voltage space.
\end{abstract}
	
\maketitle

\section{Introduction}

Latched readout is an important technique for qubit readout in quantum dots that maps a qubit state, either $\ket{1}$ or $\ket{0}$, onto a third, metastable state, called the latched state~\cite{Collard2018,Studenikin2012,Kiyama2024}. It offers two major advantages over non-latching, spin-to-charge conversion readout schemes: first, the latched state involves a reservoir transition, inducing a change in total charge occupation, and it thus generates a larger, more easily measured readout signal~\cite{Collard2019,Fogarty2018,Urdampilleta2019,Connors2020,Jang2021,Ma2024,Kojima2021,Geng2025,tsoukalas2025}. Second, the lifetime of the latched state can be controlled by tuning the tunnel rates of the system. This feature is especially useful for systems in which the relaxation rate of the qubit is much faster than the readout bandwidth of the measurement circuit \cite{Corrigan2023,Hendrickx2021,Marton2023,Song2024,Nakajima2017,kelly2025}. 

Latched states with a lifetime sufficient for measurement, however, will have a slow natural reset rate that inherently leads to long qubit re-initialization times, as described in \cite{park2024}. Long initialization times not only drastically increase measurement times in qubit experiments, but they can also be highly detrimental to quantum error correction (QEC).
For example, a measurement error rate budget of $0.5\%$ has been proposed for practical implementations of QEC \cite{Martinis2015}. To achieve this threshold for both measurement and initialization without an active reset, the latch natural decay must be slow compared to the measurement time, making the re-initialization time orders of magnitude longer than the minimum measurement time.
This is not desirable because all other qubits waiting during this time will be prone to idling errors. An active reset would mitigate these errors by re-initializing the qubit out of the latched state quickly, on-demand, and with high fidelity. Further, since readout and idling errors contribute to the overall error budget, reducing those errors eases those requirements for other contributions to the error budget, such as gate errors \cite{Hetenyi2024}.

In this work we demonstrate on-demand re-initialization of a quantum dot hybrid qubit (QDHQ) \cite{Shi2012} out of a latched state using only baseband pulses. The active reset works by pulsing to a region in voltage space where the latched state quickly decays to the qubit ground state through a fast two-step process. 
The latch we employ is the 4-electron (3,1) state, which differs by an entire electron charge from the non-latched readout states (4,1) and (3,2). The active reset method proceeds in two steps: first, the (3,1) latched state quickly decays to a (3,2) charge state via a reservoir transition at a rate of $\Gamma _R$, which is tuned to be fast. Then, an inelastic tunneling event brings the (3,2) charge state of the qubit to $(4,1)_g$, the qubit $\ket{0}$, completing the qubit re-initialization. This step occurs at a rate of $1/T_1$, which can also be tuned to be fast via the interdot barrier gate. We demonstrate that this procedure produces on-demand re-initialization of the qubit at a rate that is more than 50 times faster than the natural decay of the latched state, achieving $>99.5\%$ initialization fidelity within $4 \, \mu$s. We also describe the physical processes in the surrounding neighborhood of the optimal latch reset region, and demonstrate why those regions should be avoided for fast and effective reset.

\section{Latched Readout}

\begin{figure}[t]
\includegraphics[width = \columnwidth]{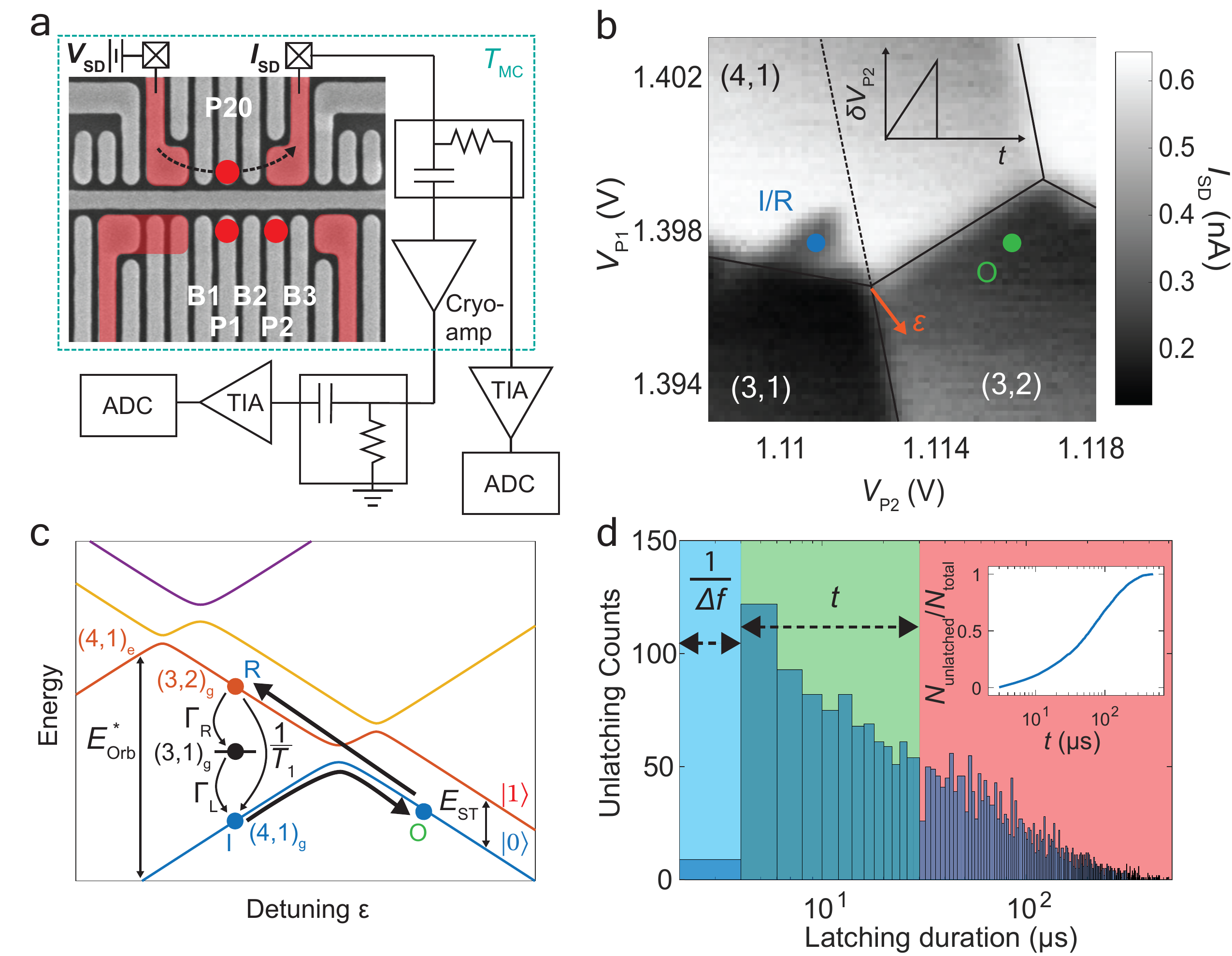}
  \caption{ 
(a) SEM micrograph of a Tunnel Falls triple dot device lithographically identical to the one used in this experiment. The circuit shown is used both for time-averaged measurements and fast, single-shot measurements. For the latter, a cryogenic amplifier is mounted at the mixing chamber (MC) stage before the transimpedance amplifier (TIA) and the analog-to-digital converter (ADC), the latter of which both are at room temperature. (b) Stability diagram with an applied excitation pulse (inset) in the (4,1)-(3,2) charge configuration with tunnel rates tuned to latch onto a (3,1) charge state. The excitation pulse amplitude takes the system from point I/R (initialization/readout) to point O (operation). (c) Energy dispersion diagram of the QDHQ in the (4,1)-(3,2) charge regime with the latched state shown along with its tunable parameters. (d) A distribution of the lifetimes of many latching events showing the inherently long initialization times that come with long lived latched states. The inset reports the cumulative distribution of the latch durations as the fraction of latching events that have unlatched, $N_\text{unlatched}/N_\text{total}$, as a function of time $t$, demonstrating that a non-active reset requires waiting much longer than the minimum measurement time in order for re-initialization to occur.}
\label{fgr:QDHQ_latch}
\end{figure}

Fig.~\ref{fgr:QDHQ_latch}a shows a scanning electron micrograph (SEM) image of an Intel Tunnel Falls Si/SiGe quantum device \cite{Neyens2024,George2024} nominally identical to the one studied here. It hosts three quantum dots and a charge sensor. A double dot is formed under plunger gates P1 and P2, and charge sensing is performed with the single-electron transistor (SET) formed under P20. Reservoirs for each dot are shown by red false coloring, with the finger gates of the third dot in the qubit channel acting as an extended reservoir. The SET response is measured through a bias tee, allowing both for slow, time-averaged measurements and for fast, single-shot measurements through a cryogenic amplifier thermalized to the mixing chamber plate.

Fig.~\ref{fgr:QDHQ_latch}b reports a stability diagram showing the $(N_{P1},N_{P2})\rightarrow$ (4,1)-(3,2) charge configuration, which hosts the $\ket{0}$ and $\ket{1}$ states of the 5-electron QDHQ, as shown in Fig.~\ref{fgr:QDHQ_latch}c. The 2-electron energy splitting of the right dot, $E_\text{ST}$, is a valley-like singlet-triplet splitting. In contrast, because there are two additional electrons in the left dot, which do not participate in the qubit dynamics, the lowest 4-electron energy splitting of the left dot is an orbital-like singlet-triplet splitting $E_\text{orb}^*$ \cite{Higginbotham2014}.
In order to enter the (3,1) latched state from the $(3,2)_g$ state in the readout window, the right dot tunnel rate $\Gamma_R$ must be tuned to be faster than the interdot relaxation rate $1/T_1$, so that an electron tunnels out of the right dot into the right lead before an inelastic tunneling event relaxes the double dot from $(3,2)$ to $(4,1)$, as shown in Fig.~\ref{fgr:QDHQ_latch}c. The lifetime of the latched state is then tunable with the left dot tunnel rate $\Gamma_L$, which controls how long it takes for an electron to load onto the left dot from the left lead, re-initializing the qubit to the $(4,1)$ ground state \cite{park2024}. For these experiments, the lifetime of the latched state was tuned to the range of $50-85\,\mu$s.

To confirm the readout is set to latch properly, we apply the voltage pulse shown in the inset of Fig.~\ref{fgr:QDHQ_latch}b. This pulse, which follows the path shown by the black arrows on the energy dispersion diagram in Fig.~\ref{fgr:QDHQ_latch}c and the points on the stability diagram in Fig.~\ref{fgr:QDHQ_latch}b $(\text{R}\rightarrow \text{O}\rightarrow \text{I})$, populates the $(3,2)_{g}$ state when the pulse abruptly returns across the polarization line. The readout window is then visible in the stability diagram shown in Fig.~\ref{fgr:QDHQ_latch}b as the dark triangular region labeled I/R (initialization/readout). It appears as a darker shade of gray, closer to that of the (3,1) region, because the latching state (3,1) is occupied for a significant fraction of the time in the readout window.


The natural reset out of the latched state is very slow, which we now demonstrate by applying the same pulse shown in the inset of Fig.~\ref{fgr:QDHQ_latch}b at point I/R and monitoring the charge sensor current using the cryogenic amplifier. We extract both latching events and unlatching events using single-shot readout (see Appendix A for measurement details).
Fig.~\ref{fgr:QDHQ_latch}d shows the resulting distribution of lifetimes for $\sim 3000$ latching events with a characteristic lifetime of $85\,\mu$s. The blue region corresponds to duration times less than or of order the readout circuit bandwidth, $1/\Delta f$, where few unlatching events can be observed. The unlatching counts contained in the green region corresponds to the number of latching events that successfully unlatched, $N_{\text{unlatched}}$, up to a time $t$. The unlatching counts contained in the red region then correspond to the number of latching events that failed to unlatch within $t$. We plot in the inset the cumulative number of unlatching events up to a time $t$, making it clear that $t$ must get very large before the cumulative unlatching distribution plateaus.
This long tail of unlatching events lasting hundreds of microseconds, visible in both the main panel and the cumulative distribution shown in the inset, motivates the need for an on-demand reset out of the latched state.

\begin{figure}[!b]
\includegraphics[width = \columnwidth]{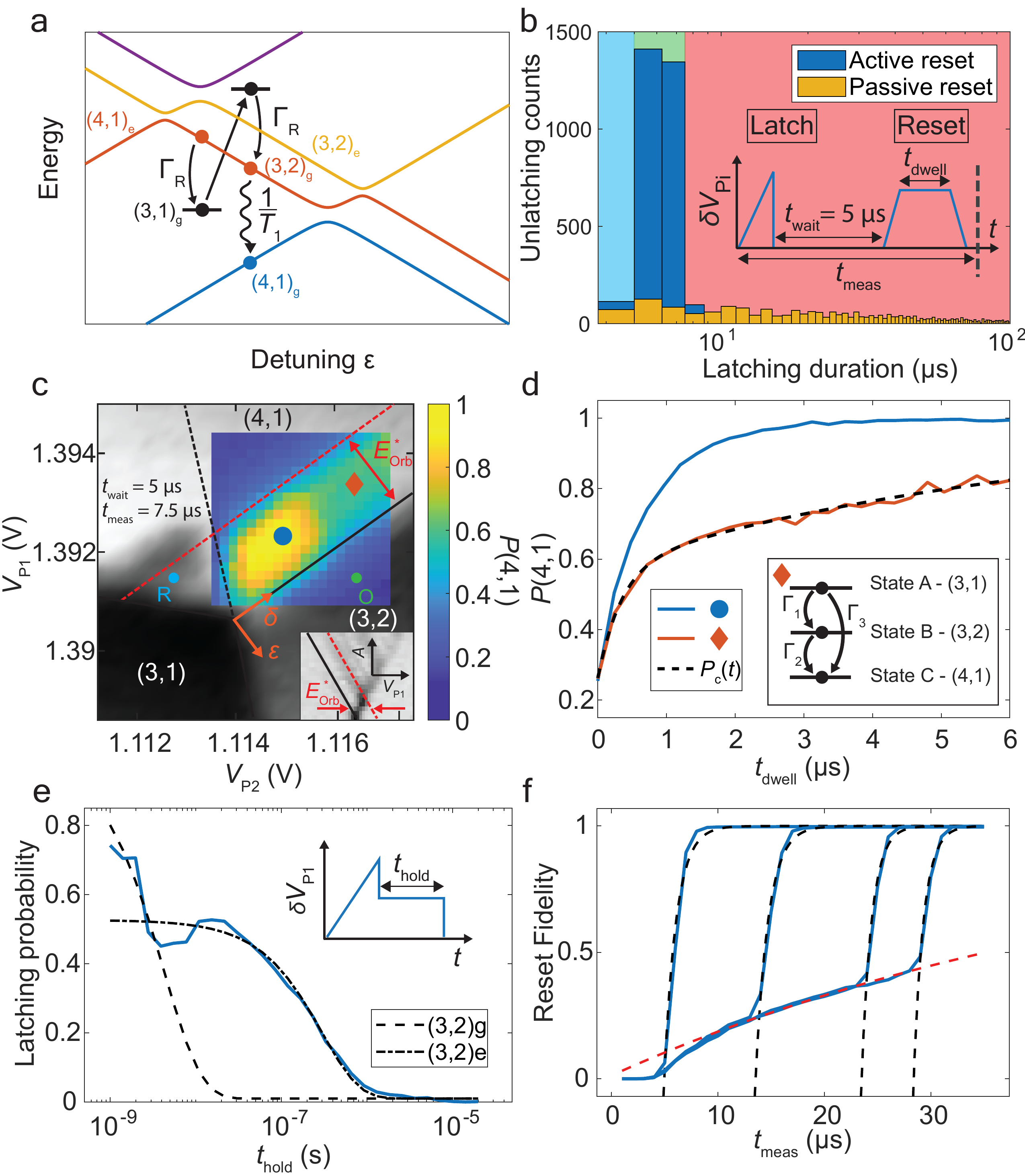}
  \caption{  
(a) QDHQ dispersion diagram showing how the reload pulse, shown in the inset of (b), initializes the qubit quickly out of the latched state in a fast two-step process. (b) Histogram showing the lifetimes of $\sim3000$ latching events with (orange) and without (blue) the active reset pulse shown in the inset being applied. (c) Baseband reset data overlaid on top of a stability diagram showing the region where the initialization happens the best. This region is bounded in detuning by $E_\text{orb}^*$ which was measured using pulsed gate spectroscopy (inset). Two distinct regions are indicated by the blue dot and red diamond, where an abrupt change in reset fidelity is observed. (d) Reset fidelity as a function of $t_\text{dwell}$ in both regions indicated in (c). Inset shows the model used to fit to the data in the red diamond region. (e) $T_1$ data measured in the blue dot region utilizing the pulse used in the inset. Two different curves are observed which are attributed to relaxation from either $(3,2)_g$ or $(3,2)_e$ to $(4,1)_g$. (f) Reset fidelity as a function of $t_\text{meas}$ taken for four different values of the $t_\text{wait}$ pulse parameter. The rate of change of the reset fidelity follows that of the natural decay of the latched state (fit to the red dashed line) until the reload pulse is sent at which point the rate of change significantly increases (black dashed line), showing that we can re-initialize our qubit on-demand, quickly, and with high fidelity.
}
\label{fgr:reset}
\end{figure}

\section{On-Demand, Active Reset of Latched Readout}

We now show that a baseband voltage pulse provides an on-demand reset by positioning the qubit at an appropriate location in gate-voltage space. Fig.~\ref{fgr:reset}a illustrates the energy level manipulations for the active reset process corresponding to the gate voltage pulse shown in the inset of Fig.~\ref{fgr:reset}b. After applying the latching pulse and waiting $t_\text{wait}=5\,\mu$s, long enough to observe the latching signal, we apply an active reset pulse that ramps the latched state to a point in voltage space where rapid unlatching occurs. By waiting at that location for a time $t_\text{dwell}$, the latched state resets far more quickly than the natural latch lifetime.

To identify the proper location for active reset, we plot in Fig.~\ref{fgr:reset}c the probability $P(4,1)$ of returning to a (4,1) state as a function of the position in gate voltage space of the reset pulse. For reference, this color plot is superimposed on a gray scale plot showing a stability diagram acquired in the presence of the latching pulse from Fig.~\ref{fgr:QDHQ_latch}b with $t_\text{dwell}=2\,\mu$s and $t_\text{meas}=7.5\,\mu$s, with the readout point shown in light blue and labeled R. $P(4,1)$ is clearly maximized near the dark blue point, corresponding to the optimum reset point.

Figure~\ref{fgr:reset}b reports in blue the unlatching counts for measurements including the active reset pulse at the location of the dark blue circle in Fig.~\ref{fgr:reset}c, revealing a very rapid reset out of the latch. For comparison, the corresponding distribution without the active reset pulse is also shown and labeled as a passive reset, revealing the expected slower re-initialization from the natural decay of the latched state.

To demonstrate that the active reset can achieve high fidelity, we report as the blue curve in Fig.~\ref{fgr:reset}d the probability $P(4,1)$ as a function of the time $t_\text{dwell}$ that the system sits at the optimal active reset point. This curve represents the reset fidelity for $t_\text{meas}=11.5\,\mu s$, which rises to $99.5\%$ within $4\ \mu s$.

We now describe the mechanism for the active reset. The active reset in the high-fidelity region occurs through the two-step process shown in Fig.~\ref{fgr:reset}a. This high-fidelity region is bounded on the left by the extended P2 transition line, shown as the black dashed line in Fig.~\ref{fgr:reset}c. To the right of that line, the (3,1) latching state is higher in energy than $(3,2)_g$ and $(3,2)_e$, enabling an electron to load quickly onto the P2 dot from the right reservoir at a rate of $\Gamma_R$, which has already been tuned to be fast, as required for latched readout. In this device, the right dot ST splitting $E_\text{ST}<16\,\mu$eV, is quite small, making both the $(3,2)_g$ and $(3,2)_e$ states energetically accessible. After the system transitions out of the (3,1) latching state into (3,2), an inelastic tunneling event returns the system to $(4,1)_g$, completing the re-initialization through this $T_1$ decay.

The speed of this interdot decay process is important, and for that reason we now measure the $T_1$ time using the pulse shown in the inset of Fig.~\ref{fgr:reset}e. The pulse consists of an adiabatic ramp across the polarization line between the points labeled R (readout) and O (operation) in Fig.~\ref{fgr:reset}c, occupying a mixture of $(3,2)_g$ and $(3,2)_e$ because of the small $E_\text{ST}$. The state then abruptly pulses to the region of high reset fidelity where the latched state is not accessible and is held for a time $t_\text{hold}$ to allow either (3,2) state to relax to $(4,1)_g$. After a time $t_\text{hold}$, the pulse abruptly returns to the readout point where the latched state is accessible to the excited state. Plotted in Fig.~\ref{fgr:reset}e is the latching probability, which decreases as a function of $t_\text{hold}$ as more states relax to $(4,1)_g$ through the $T_1$ process. Two exponential curves are visible in the data. We argue that these correspond to two relaxation processes to $(4,1)_g$ from $(3,2)_g$ and $(3,2)_e$, respectively. The extracted $T_1$ times are $301$~ns for $(3,2)_e\rightarrow(4,1)_g$, and $4.3$~ns for $(3,2)_g\rightarrow(4,1)_g$. We assign the slower transition rate to the $(3,2)_e\rightarrow(4,1)_g$ process, because we measure the excited state tunnel coupling between $(3,2)_e$ and $(4,1)_g$ to be smaller than the ground state tunnel coupling between $(3,2)_g$ and $(4,1)_g$ (see Appendix B for details).

To demonstrate that the active reset functions on-demand, Fig.~\ref{fgr:reset}f reports the reset fidelity as a function of $t_\text{meas}$ acquired for four different values of the reset pulse parameter $t_\text{wait}$. The rate of change of the reset fidelity follows that of the natural decay of the latched state (slowly increasing curve fit with the red dashed line), until the reload pulse is sent at time $t_\text{wait}$, at which point the rate of change significantly increases (four curves fit with black dashed lines), showing that with this active reset pulse, the qubit begins a rapid reset as soon as the active reset pulse is applied.

Having discussed how and why the active reset works in the region indicated by the blue dot in Fig.~\ref{fgr:reset}c, we now look at what happens outside this region in voltage space. 
We look first at a region of similar double dot detuning $\varepsilon$ but larger $\delta$ (the direction perpendicular to the $\varepsilon$ axis), as indicated by the red diamond in Fig.~\ref{fgr:reset}c, which marks a location in voltage space where there is a significant decrease in $P(4,1)$. In this regime, the latched state (3,1) is higher in energy, so that additional (3,2) excited states are accessible for the latched state to decay into, and these states can be longer lived. Performing the same $t_\text{dwell}$ measurement as before in this region, the red line in Fig.~\ref{fgr:reset}d initially begins to increase rapidly, but after $\sim 1\ \mu s$, the rate of change of $P(4,1)$ drastically slows down. 

This behavior suggests that the system becomes ``stuck" in a long-lived (3,2) charge state when reseting at this location. We model this behavior using the three state model shown in the inset of Fig.~\ref{fgr:reset}d, where state A is the (3,1) latched state, state B is a higher lying (3,2) charge state, and state C is a (4,1) charge state. Assuming exponential decays as shown in the inset of Fig.~\ref{fgr:reset}d with decay rates given by $\Gamma_1,\ \Gamma_2,$ and $\Gamma_3$, and an initial condition of starting in state A, we solve a set of three coupled rate equations, similar to the model described in the supplemental material in Ref.~\cite{Simmons2011}, that yield a probability of ending up in state C given by,
\begin{align*}
    P_c(t) =& \frac{\Gamma_1}{\Gamma_1-\Gamma_2+\Gamma_3}(1-e^{-\Gamma_2t})\\
    &+\frac{\Gamma_3(\Gamma_1-\Gamma_2+\Gamma_3)-\Gamma_1\Gamma_2}{(\Gamma_1-\Gamma_2+\Gamma_3)(\Gamma_1+\Gamma_3)}(1-e^{-(\Gamma_1+\Gamma_3)t}).
\end{align*}
The dashed black line in Fig.~\ref{fgr:reset}d
is a fit of the data to this expression, and it captures well the shape of the red curve. The extracted decay rates from this fit reveal that $1/\Gamma_2=6.87\ \mu$s is approximately 10 times slower than $1/\Gamma_1=0.64\ \mu$s and $1/\Gamma_3=0.75\ \mu$s.  These parameters are consistent with the hypothesis that there is a long-lived (3,2) excited state accessible at the location of the red diamond in Fig.~\ref{fgr:reset}c, and that the effect of that state is visible in the slow rise of the red curve in Fig.~\ref{fgr:reset}d. More generally, we note that in a double dot with 5 electrons there are other states that could also be long lived, and in Appendix C we describe one such example.

\begin{figure}[h]
\includegraphics[width=\columnwidth]{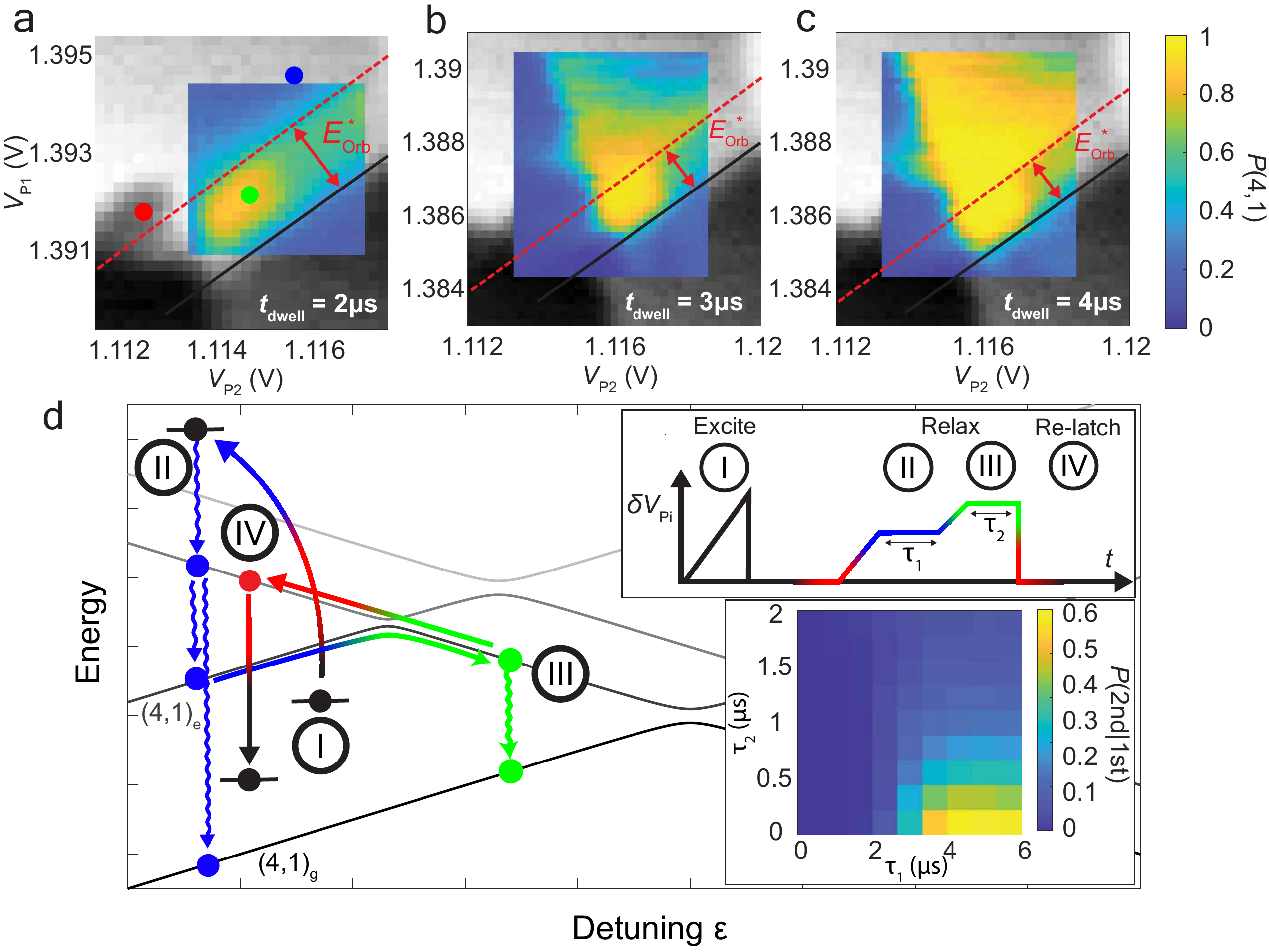}
  \caption{ 
Pulsed-gate measurements to understand the decay pathways for $\varepsilon > E_\text{orb}^{*}$. (a)-(c) The probability of transitioning from the latched state to a (4,1) state using the pulse shown in the inset of Fig.~\ref{fgr:reset}b for three different $t_\text{dwell}$ times. The increase in $P(4,1)$ for $\varepsilon > E_\text{orb}^{*}$ includes transitions to state $(4,1)_e$. (d) Dispersion diagram showing how the pulse shown in the upper inset can differentiate between the decay paths to $(4,1)_e$ and $(4,1)_g$. The ramp between stages II and III maps the $(4,1)_g$ state to itself, resulting in only one latching signal. That ramp maps the $(4,1)_e$ state to a $(3,2)$ state, which will then relatch, resulting in an observable second latching event. The probability of seeing a second latching event given that a first latching event, $P$(2nd$|$1st), is shown in the lower inset as a function of the dwell times for each pulse, $\tau_1$ and $\tau_2$. $P$(2nd$|$1st) reaches a value as large as ~60\%, indicating that the latched state often decays to $(4,1)_e$ when $\varepsilon > E_\text{orb}^{*}$.}

\label{fgr:double_dwell}
\end{figure}

Moving to the upper left from the blue point in Fig.~\ref{fgr:reset}c, the high-fidelity reset region stops at the location of the dashed red line, which corresponds to the singlet-triplet splitting of the left dot, $E_\text{orb}^*$, as determined by the pulsed gate spectroscopy measurements shown in the lower inset. It is interesting to look at what happens when attempting to reset at a location to the upper left of this red line, because the behavior is qualitatively different in that regime. Figs.~\ref{fgr:double_dwell}a-c show again a 2D map of the probability $P(4,1)$ acquired using the pulse shown in the inset of Fig.~\ref{fgr:reset}b, acquired now for three values of $t_\text{dwell}$.
Focusing on the region to the upper left of the dashed red line, we note that in this region the (3,2) states, easily accessible from the (3,1) latched state, can decay in turn either to $(4,1)_g$ (the desired re-initialization state) or to $(4,1)_e$, as shown by the blue decay paths in Fig.~\ref{fgr:double_dwell}d. 

Because $(4,1)_g$ and $(4,1)_e$ produce the same charge sensing signal, to differentiate between them we employ a special pulse sequence that traverses the various energy states and anti-crossings in the dispersion diagram, as shown in Fig.~\ref{fgr:double_dwell}d. Step I initiates the (3,1) latch; Step II offers decays to either $(4,1)_g$ or $(4,1)_e$; 
to differentiate between these two, the voltage ramp to Step III shifts the states in detuning, mapping $(4,1)_g$ to itself (which does nothing), and mapping $(4,1)_e$ to the (3,2) state, which then relatches at Step IV. Thus, by employing this pulse sequence and looking for a second latching event, we can identify which (4,1) state was occupied and with what probability.

The lower right inset of Fig.~\ref{fgr:double_dwell}d shows the probability of seeing a second latching event given the observation of the first latching event, $P$(2nd$|$1st). The bright yellow region indicates a significant probability of observing such a second latching event. In this region, the latched state (3,1) decays to $(4,1)_e$, which then maps adiabatically to a (3,2) state during the transition to Step III. For short $\tau_2$, where little time is permitted for the qubit at point III to relax to $(4,1)_g$, the system transitions to the red point, where during Step IV it relatches. For longer $\tau_2$, the system decays to $(4,1)_g$ before leaving Step III. This measurement confirms that to the upper left of the dashed red line in Figs.~\ref{fgr:double_dwell}a-c the latched state decays frequently to $(4,1)_e$, and therefore this region should be avoided when choosing an active reset location.

\section{Discussion}
To ameliorate the issue of long initialization times out of a latched state, we presented an active baseband reset pulse that re-initializes a qubit on-demand, quickly, and with high fidelity. The efficiency of this reset relies on the fast inelastic tunneling process that occurs in the charge-like regime of a double quantum dot. The window for this reset is bounded by the ST splitting, $E_{orb}^*$, and corresponds to the region of high reset fidelity shown in Fig.~\ref{fgr:reset}c. In this window, we are able to re-initialize our qubit out of a latched state on-demand, with a reset fidelity $>99.5\%$ within $4\ \mu$s, significantly faster than relying on the natural decay of the latched state while maintaining excellent latched readout.

In addition to the regions of high reset fidelity, we also identified regions to avoid when using the active reset pulse. Within the ST splitting of the qubit, when attempting to initialize in regions of larger $\delta$, like in the red diamond region in Fig.~\ref{fgr:reset}c, additional excited states become energetically accessible to the latched state. These additional excited states can have significantly longer relaxation times, resulting in a reduction of the reset fidelity. We considered a simple three state model, which fits the data in this region quite well.
The region at larger detuning, beyond the ST splitting, should also be avoided, because it enables relaxation out of the latch to $(4,1)_e$, which is not the desired re-initialization state.

The active reset out of a latched state in this work was performed on a QDHQ in Si/SiGe; however, this technique can be applied to other types of spin qubits as well. Latched Pauli spin blockade (L-PSB) is a common readout method employed in other types of spin qubits, such as singlet-triplet qubits, exchange-only qubits, and even Loss-DiVincenzo qubits for parity readout \cite{Hutin2019,Bogan2019,Vahapoglu2021}. L-PSB latches standard PSB states onto a third metastable state by pulsing into the L-PSB window with the proper tuning of reservoir tunnel rates, analogous to how latched readout was performed here. The same two-step decay process discussed here can be applied to L-PSB, by pulsing back into the standard PSB window. In this window the latched state first quickly decays to the excited triplet state $\ket{T_0(1,1)}$ of the qubit through a fast reservoir transition, then an inelastic tunneling event brings the qubit to the singlet ground state, $\ket{S(2,0)}$, completing the initialization. This second step is the key component to having this active reset work quickly. In the QDHQ, we take advantage of the additional charge degree of freedom at low double dot detunings to allow the qubit to reach its ground state quickly. For L-PSB, the speed of this step will be set by the singlet-triplet relaxation rate. In materials with strong spin-orbit coupling, such as germanium, the singlet-triplet relaxation rate is typically on the order of microseconds, which would allow the qubit to initialize relatively quickly. In materials with weak spin-orbit coupling, like silicon, the singlet-triplet relaxation rate tends to be much longer, on the order of tens of milliseconds, which would significantly slow the rate of initialization. This technique could, however, be used in conjunction with other methods of increased singlet-triplet relaxation to speed up qubit initialization, such as the $T_1$ ``hot spots" around ST qubit anti-crossings, as demonstrated in Ref.~\cite{Blumoff2022}.

\section*{Acknowledgments}

The authors thank HRL Laboratories for support and Joe Kerckhoff for valuable discussions. This research was sponsored in part by the Army Research Office under Award No.\ W911NF-23-1-0110 and under Cooperative Agreement No.\ W911NF-22-2-0037. This material is based upon work supported by the U.S. Department of Energy Office of Science National Quantum Information Science Research Centers as part of the Q-NEXT center, which was important for initial tuning of the Tunnel Falls device.
The views, conclusions, and recommendations contained in this document are those of the authors and are not necessarily endorsed by nor should they be interpreted as representing the official policies, either expressed or implied, of the Army Research Office or the U.S. Government. The U.S. Government is authorized to reproduce and distribute reprints for U.S. Government purposes notwithstanding any copyright notation herein.

\section*{Appendix A: Single-shot data analysis}

\begin{figure}[h]
\includegraphics[width = \columnwidth]{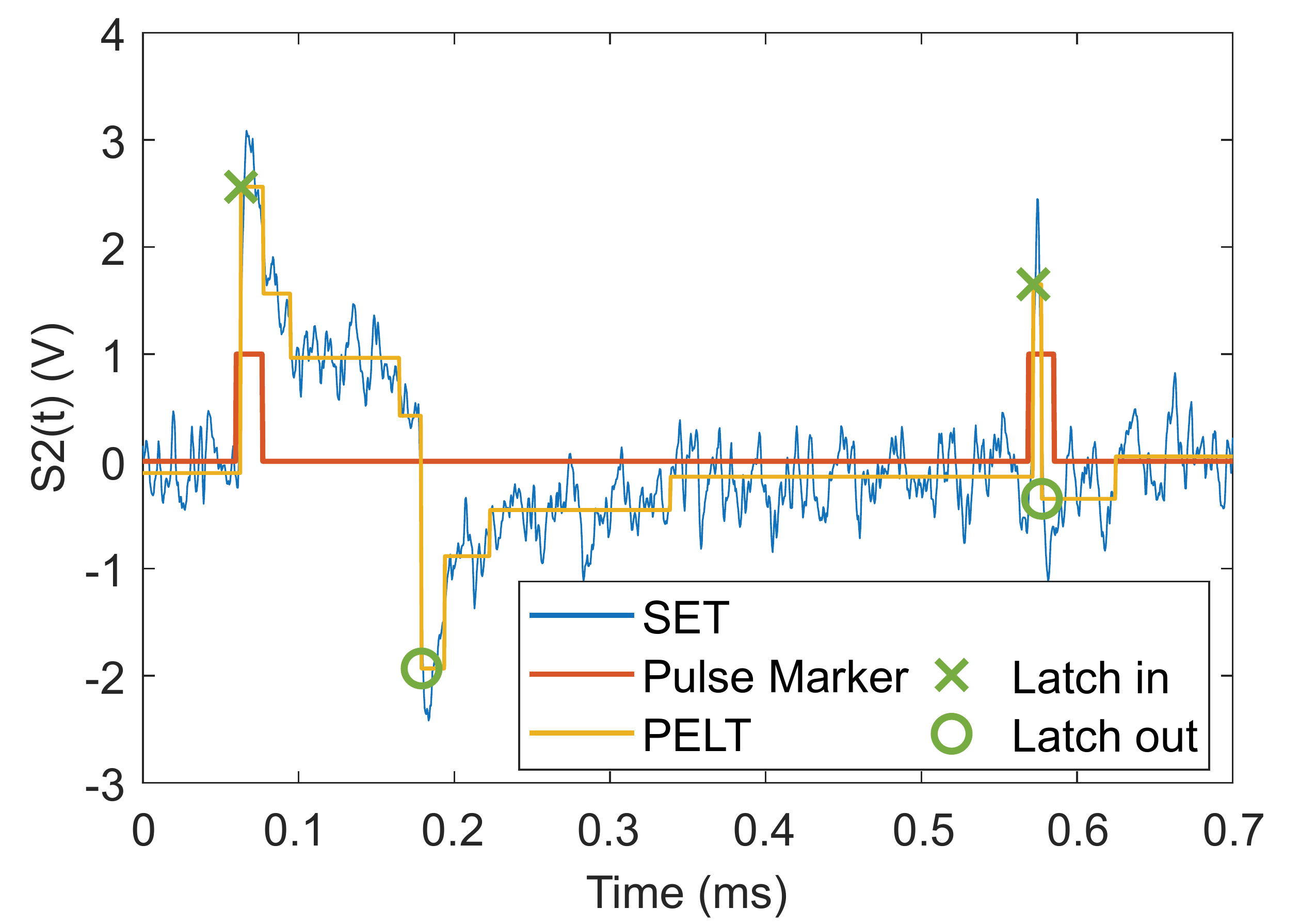}
  \caption{  
Example single-shot data used in this work. The blue trace is the raw SET data and the yellow trace is the transformed PELT data showing the abrupt changes in the average value of the SET data. The red trace is the pulse marker and shows when a pulse is applied. Visible are two example latching events, one that is long lived whose signal response is affected by the high pass filter, and a short lived latching event whose signal response isn't affected by the high pass filter.
}
\label{fgr:time_trace}
\end{figure}

The bias tee after the cryogenic amplifier acts as a high pass filter for the charge sensing signal. Because of this high pass effect, we use a two-part algorithm to effectively track the start and end times of latching events.
Fig.~\ref{fgr:time_trace} shows an example time trace that exhibits both parts of the algorithm. We start by applying a function to our SET data that uses the pruned exact linear time (PELT) method \cite{Killick2012} to look for abrupt changes in its average value, resulting in the yellow trace in Fig.~\ref{fgr:time_trace}. With this transformed data, we search for an initial latching event within 10~$\mu s$ of the applied pulse. Such events are cleanly identified by a sharp step up, and they can be easily timestamped. For long-lived latching events, the high pass filter returns the SET response to 0~V before an unlatching event, as is visible in the first event in Fig.~\ref{fgr:time_trace}. In this case, when the system unlatches, the PELT trace shows a clear step down, which again is easy to timestamp. Short-lived latching events, on the other hand, are not affected by the high pass effect, and therefore the end of the latching event cannot be timestamped by looking for a negative peak: the downward step does not go very negative, because it is starting from a positive value. Instead, when no negative steps are found, the algorithm looks for large negative changes in the PELT data close to the start of the latching event, which correspond to the unlatching event of a short-lived latched state, as is visible for the second event in Fig.~\ref{fgr:time_trace}.

\section*{Appendix B: Tunnel coupling measurements}
\begin{figure}[b]
\includegraphics[width = \columnwidth]{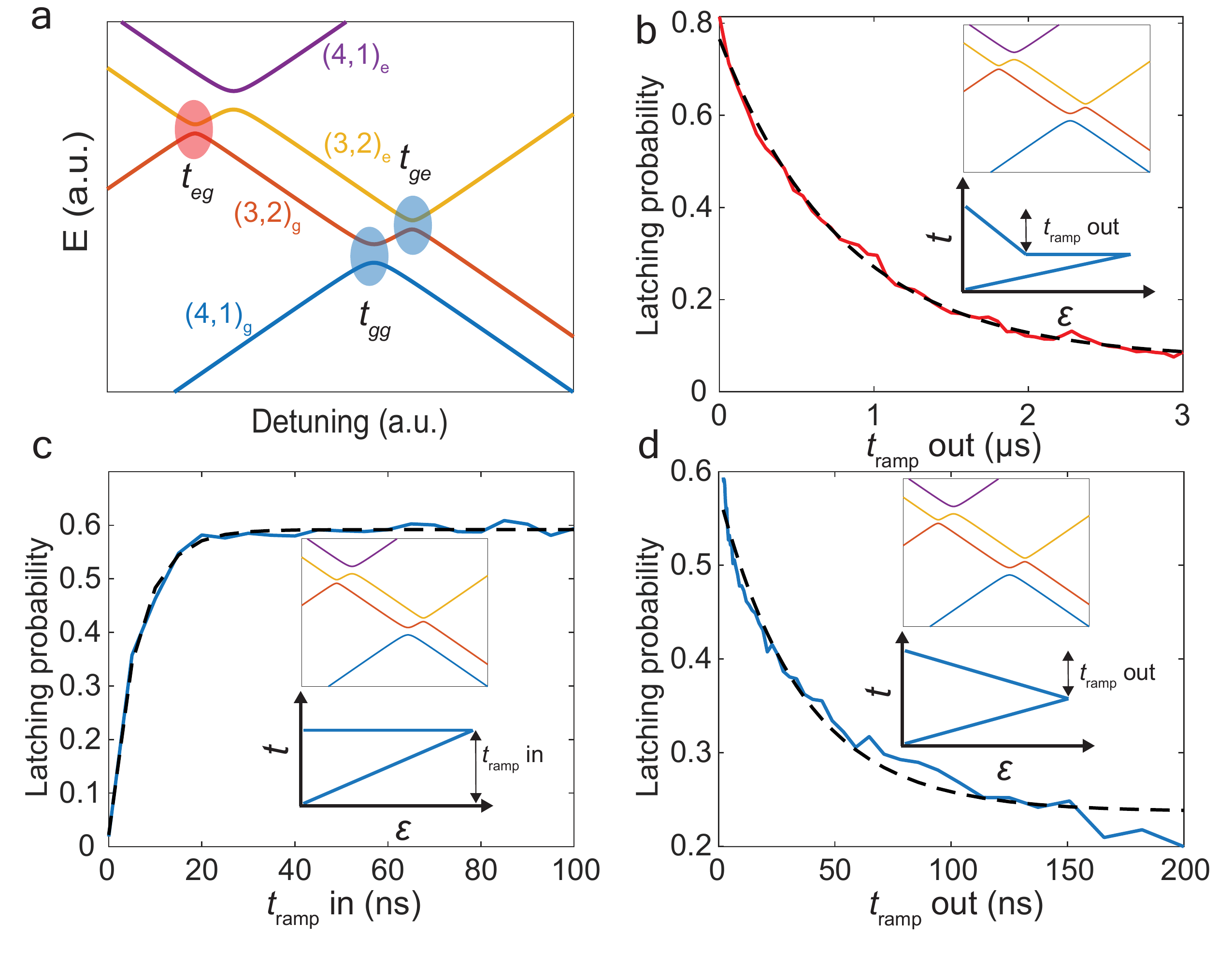}
  \caption{  
Ground and excited state tunnel coupling measurements. (a) Dispersion diagram showing which tunnel couplings are extracted. $t_{eg}$, highlighted in red, is extracted from the data in (b), and $t_{gg}$ and $t_{ge}$, highlighted in blue, are extracted by simultaneous fitting to (c) and (d). (b). $t_{eg}$ data acquired using the pulse shown in the inset. The pulse is tuned to abruptly move across the ground state anti-crossing to enter an excited state via a LZ transition. Then the ramp time across the $t_{eg}$ anti-crossing is swept. (c-d). $t_{gg}$ and $t_{ge}$ data acquired using the pulse shown in the inset. Both tunnel couplings are extracted by simultaneous fitting to both data sets.
}
\label{fgr:tunnel_coupling}
\end{figure}

We measure the ground and excited state tunnel couplings of the double dot by inducing Landau-Zener transitions and measuring the probability of observing a latching event as a function of the ramp times across the anticrossings, as shown in Fig.~\ref{fgr:tunnel_coupling}. The excited state tunnel coupling $t_{eg}$ is extracted by fitting an exponential to the data in Fig.~\ref{fgr:tunnel_coupling}b, and it is found to be $t_{eg}=31$ MHz, much smaller than the values for $t_{gg}$ and $t_{ge}$ that we report below. 

Due to a small valley splitting in this device, it becomes difficult to ramp across the ground state anti-crossing $t_{gg}$ without also moving across the excited state anti-crossing $t_{ge}$. For this reason, we perform two experiments, as shown in Figs.~\ref{fgr:tunnel_coupling}c,d, in order to gain enough information to extract both tunnel couplings. Starting in $(4,1)_g$, we calculate the probability of returning to state $(4,1)_g$ after ramping in and out across all four anti-crossings. We employ a four-state model, taking into account all possible Landau-Zener transitions during the pulse. The complement of this expression is the probability of ending up in any excited state, which have the ability to enter a latched state. We then simplify this expression by taking the two extreme cases of an abrupt ramp out (corresponding to the data in Fig.~\ref{fgr:tunnel_coupling}c), and an adiabatic ramp in (corresponding to the data in Fig.~\ref{fgr:tunnel_coupling}d), resulting in
\begin{equation}
    f_1(t) = 1-\exp{\Big(\frac{-2\pi}{\hbar\alpha\Delta V }(t_{gg}^2+t_{ge}^2)t_\text{ramp}\Big)}
    \label{abrupt_out}
\end{equation}
and
\begin{equation}
    f_2(t) = \exp{\Big(\frac{-2\pi}{\hbar\alpha\Delta V }t_{gg}^2t_\text{ramp}\Big)},\\
    \label{adiabatic_in}
    \vspace{0.4cm}
\end{equation}
respectively.
The data in Fig.~\ref{fgr:tunnel_coupling}c and d is then simultaneously fit to equations \ref{abrupt_out} and \ref{adiabatic_in} to extract the ground state tunnel coupling $t_{gg}=517$ MHz and the excited state tunnel coupling $t_{ge}=232$ MHz.

\section*{Appendix C: 5-electron excited states involving a right dot valley excitation}

\begin{figure}[t]
\includegraphics[width = \columnwidth]{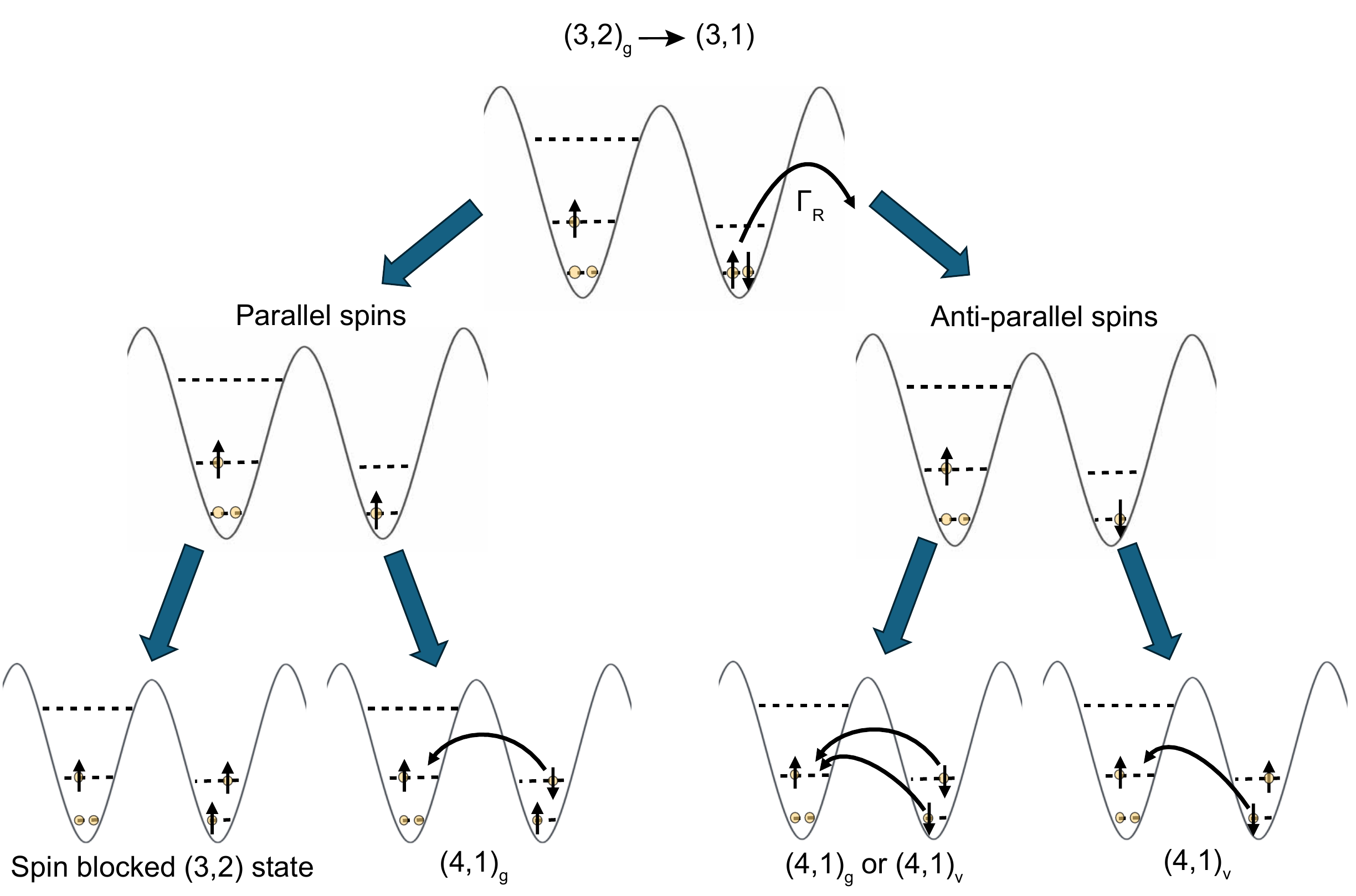}
  \caption{  
Possible spin configurations when resetting a $(3,1)$ latched state through the excited valley state of the right dot. When first entering the latched state, the two remaining valence electrons can have either parallel or anti-parallel spins (row 2). When resetting through the excited valley, either a spin up or spin down electron can tunnel onto the right dot, leading to the four possible states shown in the bottom row.
Here, the black arrows show the available (4,1) reset paths, with a spin-blocked configuration on the far left.
}
\label{fgr:leakage_state}
\end{figure}

There are many 5-electron double dot excited states that are not shown in the figures in the main text. Here, in the bottom row of Fig.~\ref{fgr:leakage_state}, we illustrate four such states that have a valley excitation in the right dot. (Without loss of generality, in this figure we assume the unpaired spin in the left dot has $S_\text{z} = +1/2$.) While during qubit operation these states are not expected to significantly couple to the four 5-electron double dot states shown in the main text, and therefore should not impact the coherent manipulation of the hybrid qubit, they 
could in principle be accessed when reloading from (3,1) to (3,2) at the location of the blue dot in Fig.~2(c). Since the electron loading from the reservoir can be either spin up or spin down, the pathways shown result in four different spin configurations of the double dot system. The state shown on the lower left is spin blockaded, and it would cause a delay in the reset to (4,1).  
We do not observe such behavior here, suggesting that the excited valley state is not accessed in our experiments. However, these interesting states could be the subject of future study.


\section*{Data-Availability}
The data that support the findings of this study are openly available in a Zenodo repository \cite{Marciniec_zenodo}.

\bibliography{ref.bib}

\begin{thebibliography}{32}%
\makeatletter
\providecommand \@ifxundefined [1]{%
 \@ifx{#1\undefined}
}%
\providecommand \@ifnum [1]{%
 \ifnum #1\expandafter \@firstoftwo
 \else \expandafter \@secondoftwo
 \fi
}%
\providecommand \@ifx [1]{%
 \ifx #1\expandafter \@firstoftwo
 \else \expandafter \@secondoftwo
 \fi
}%
\providecommand \natexlab [1]{#1}%
\providecommand \enquote  [1]{``#1''}%
\providecommand \bibnamefont  [1]{#1}%
\providecommand \bibfnamefont [1]{#1}%
\providecommand \citenamefont [1]{#1}%
\providecommand \href@noop [0]{\@secondoftwo}%
\providecommand \href [0]{\begingroup \@sanitize@url \@href}%
\providecommand \@href[1]{\@@startlink{#1}\@@href}%
\providecommand \@@href[1]{\endgroup#1\@@endlink}%
\providecommand \@sanitize@url [0]{\catcode `\\12\catcode `\$12\catcode `\&12\catcode `\#12\catcode `\^12\catcode `\_12\catcode `\%12\relax}%
\providecommand \@@startlink[1]{}%
\providecommand \@@endlink[0]{}%
\providecommand \url  [0]{\begingroup\@sanitize@url \@url }%
\providecommand \@url [1]{\endgroup\@href {#1}{\urlprefix }}%
\providecommand \urlprefix  [0]{URL }%
\providecommand \Eprint [0]{\href }%
\providecommand \doibase [0]{https://doi.org/}%
\providecommand \selectlanguage [0]{\@gobble}%
\providecommand \bibinfo  [0]{\@secondoftwo}%
\providecommand \bibfield  [0]{\@secondoftwo}%
\providecommand \translation [1]{[#1]}%
\providecommand \BibitemOpen [0]{}%
\providecommand \bibitemStop [0]{}%
\providecommand \bibitemNoStop [0]{.\EOS\space}%
\providecommand \EOS [0]{\spacefactor3000\relax}%
\providecommand \BibitemShut  [1]{\csname bibitem#1\endcsname}%
\let\auto@bib@innerbib\@empty
\bibitem [{\citenamefont {Harvey-Collard}\ \emph {et~al.}(2018)\citenamefont {Harvey-Collard}, \citenamefont {D'Anjou}, \citenamefont {Rudolph}, \citenamefont {Jacobson}, \citenamefont {Dominguez}, \citenamefont {Ten~Eyck}, \citenamefont {Wendt}, \citenamefont {Pluym}, \citenamefont {Lilly}, \citenamefont {Coish}, \citenamefont {Pioro-Ladri\`ere},\ and\ \citenamefont {Carroll}}]{Collard2018}%
  \BibitemOpen
  \bibfield  {author} {\bibinfo {author} {\bibfnamefont {P.}~\bibnamefont {Harvey-Collard}}, \bibinfo {author} {\bibfnamefont {B.}~\bibnamefont {D'Anjou}}, \bibinfo {author} {\bibfnamefont {M.}~\bibnamefont {Rudolph}}, \bibinfo {author} {\bibfnamefont {N.~T.}\ \bibnamefont {Jacobson}}, \bibinfo {author} {\bibfnamefont {J.}~\bibnamefont {Dominguez}}, \bibinfo {author} {\bibfnamefont {G.~A.}\ \bibnamefont {Ten~Eyck}}, \bibinfo {author} {\bibfnamefont {J.~R.}\ \bibnamefont {Wendt}}, \bibinfo {author} {\bibfnamefont {T.}~\bibnamefont {Pluym}}, \bibinfo {author} {\bibfnamefont {M.~P.}\ \bibnamefont {Lilly}}, \bibinfo {author} {\bibfnamefont {W.~A.}\ \bibnamefont {Coish}}, \bibinfo {author} {\bibfnamefont {M.}~\bibnamefont {Pioro-Ladri\`ere}},\ and\ \bibinfo {author} {\bibfnamefont {M.~S.}\ \bibnamefont {Carroll}},\ }\bibfield  {title} {\bibinfo {title} {High-fidelity single-shot readout for a spin qubit via an enhanced latching mechanism},\ }\href {https://doi.org/10.1103/PhysRevX.8.021046} {\bibfield  {journal}
  {\bibinfo  {journal} {Phys. Rev. X}\ }\textbf {\bibinfo {volume} {8}},\ \bibinfo {pages} {021046} (\bibinfo {year} {2018})}\BibitemShut {NoStop}%
\bibitem [{\citenamefont {Studenikin}\ \emph {et~al.}(2012)\citenamefont {Studenikin}, \citenamefont {Thorgrimson}, \citenamefont {Aers}, \citenamefont {Kam}, \citenamefont {Zawadzki}, \citenamefont {Wasilewski}, \citenamefont {Bogan},\ and\ \citenamefont {Sachrajda}}]{Studenikin2012}%
  \BibitemOpen
  \bibfield  {author} {\bibinfo {author} {\bibfnamefont {S.~A.}\ \bibnamefont {Studenikin}}, \bibinfo {author} {\bibfnamefont {J.}~\bibnamefont {Thorgrimson}}, \bibinfo {author} {\bibfnamefont {G.~C.}\ \bibnamefont {Aers}}, \bibinfo {author} {\bibfnamefont {A.}~\bibnamefont {Kam}}, \bibinfo {author} {\bibfnamefont {P.}~\bibnamefont {Zawadzki}}, \bibinfo {author} {\bibfnamefont {Z.~R.}\ \bibnamefont {Wasilewski}}, \bibinfo {author} {\bibfnamefont {A.}~\bibnamefont {Bogan}},\ and\ \bibinfo {author} {\bibfnamefont {A.~S.}\ \bibnamefont {Sachrajda}},\ }\bibfield  {title} {\bibinfo {title} {Enhanced charge detection of spin qubit readout via an intermediate state},\ }\href {https://doi.org/10.1063/1.4749281} {\bibfield  {journal} {\bibinfo  {journal} {Applied Physics Letters}\ }\textbf {\bibinfo {volume} {101}},\ \bibinfo {pages} {233101} (\bibinfo {year} {2012})}\BibitemShut {NoStop}%
\bibitem [{\citenamefont {Kiyama}\ \emph {et~al.}(2024)\citenamefont {Kiyama}, \citenamefont {van Hien}, \citenamefont {Ludwig}, \citenamefont {Wieck},\ and\ \citenamefont {Oiwa}}]{Kiyama2024}%
  \BibitemOpen
  \bibfield  {author} {\bibinfo {author} {\bibfnamefont {H.}~\bibnamefont {Kiyama}}, \bibinfo {author} {\bibfnamefont {D.}~\bibnamefont {van Hien}}, \bibinfo {author} {\bibfnamefont {A.}~\bibnamefont {Ludwig}}, \bibinfo {author} {\bibfnamefont {A.~D.}\ \bibnamefont {Wieck}},\ and\ \bibinfo {author} {\bibfnamefont {A.}~\bibnamefont {Oiwa}},\ }\bibfield  {title} {\bibinfo {title} {High-fidelity spin readout via the double latching mechanism},\ }\href {https://doi.org/10.1038/s41534-024-00882-1} {\bibfield  {journal} {\bibinfo  {journal} {npj Quantum Information}\ }\textbf {\bibinfo {volume} {10}},\ \bibinfo {pages} {95} (\bibinfo {year} {2024})}\BibitemShut {NoStop}%
\bibitem [{\citenamefont {Harvey-Collard}\ \emph {et~al.}(2019)\citenamefont {Harvey-Collard}, \citenamefont {Jacobson}, \citenamefont {Bureau-Oxton}, \citenamefont {Jock}, \citenamefont {Srinivasa}, \citenamefont {Mounce}, \citenamefont {Ward}, \citenamefont {Anderson}, \citenamefont {Manginell}, \citenamefont {Wendt}, \citenamefont {Pluym}, \citenamefont {Lilly}, \citenamefont {Luhman}, \citenamefont {Pioro-Ladri\`ere},\ and\ \citenamefont {Carroll}}]{Collard2019}%
  \BibitemOpen
  \bibfield  {author} {\bibinfo {author} {\bibfnamefont {P.}~\bibnamefont {Harvey-Collard}}, \bibinfo {author} {\bibfnamefont {N.~T.}\ \bibnamefont {Jacobson}}, \bibinfo {author} {\bibfnamefont {C.}~\bibnamefont {Bureau-Oxton}}, \bibinfo {author} {\bibfnamefont {R.~M.}\ \bibnamefont {Jock}}, \bibinfo {author} {\bibfnamefont {V.}~\bibnamefont {Srinivasa}}, \bibinfo {author} {\bibfnamefont {A.~M.}\ \bibnamefont {Mounce}}, \bibinfo {author} {\bibfnamefont {D.~R.}\ \bibnamefont {Ward}}, \bibinfo {author} {\bibfnamefont {J.~M.}\ \bibnamefont {Anderson}}, \bibinfo {author} {\bibfnamefont {R.~P.}\ \bibnamefont {Manginell}}, \bibinfo {author} {\bibfnamefont {J.~R.}\ \bibnamefont {Wendt}}, \bibinfo {author} {\bibfnamefont {T.}~\bibnamefont {Pluym}}, \bibinfo {author} {\bibfnamefont {M.~P.}\ \bibnamefont {Lilly}}, \bibinfo {author} {\bibfnamefont {D.~R.}\ \bibnamefont {Luhman}}, \bibinfo {author} {\bibfnamefont {M.}~\bibnamefont {Pioro-Ladri\`ere}},\ and\ \bibinfo {author} {\bibfnamefont {M.~S.}\ \bibnamefont
  {Carroll}},\ }\bibfield  {title} {\bibinfo {title} {Spin-orbit interactions for singlet-triplet qubits in silicon},\ }\href {https://doi.org/10.1103/PhysRevLett.122.217702} {\bibfield  {journal} {\bibinfo  {journal} {Phys. Rev. Lett.}\ }\textbf {\bibinfo {volume} {122}},\ \bibinfo {pages} {217702} (\bibinfo {year} {2019})}\BibitemShut {NoStop}%
\bibitem [{\citenamefont {Fogarty}\ \emph {et~al.}(2018)\citenamefont {Fogarty}, \citenamefont {Chan}, \citenamefont {Hensen}, \citenamefont {Huang}, \citenamefont {Tanttu}, \citenamefont {Yang}, \citenamefont {Laucht}, \citenamefont {Veldhorst}, \citenamefont {Hudson}, \citenamefont {Itoh}, \citenamefont {Culcer}, \citenamefont {Ladd}, \citenamefont {Morello},\ and\ \citenamefont {Dzurak}}]{Fogarty2018}%
  \BibitemOpen
  \bibfield  {author} {\bibinfo {author} {\bibfnamefont {M.~A.}\ \bibnamefont {Fogarty}}, \bibinfo {author} {\bibfnamefont {K.~W.}\ \bibnamefont {Chan}}, \bibinfo {author} {\bibfnamefont {B.}~\bibnamefont {Hensen}}, \bibinfo {author} {\bibfnamefont {W.}~\bibnamefont {Huang}}, \bibinfo {author} {\bibfnamefont {T.}~\bibnamefont {Tanttu}}, \bibinfo {author} {\bibfnamefont {C.~H.}\ \bibnamefont {Yang}}, \bibinfo {author} {\bibfnamefont {A.}~\bibnamefont {Laucht}}, \bibinfo {author} {\bibfnamefont {M.}~\bibnamefont {Veldhorst}}, \bibinfo {author} {\bibfnamefont {F.~E.}\ \bibnamefont {Hudson}}, \bibinfo {author} {\bibfnamefont {K.~M.}\ \bibnamefont {Itoh}}, \bibinfo {author} {\bibfnamefont {D.}~\bibnamefont {Culcer}}, \bibinfo {author} {\bibfnamefont {T.~D.}\ \bibnamefont {Ladd}}, \bibinfo {author} {\bibfnamefont {A.}~\bibnamefont {Morello}},\ and\ \bibinfo {author} {\bibfnamefont {A.~S.}\ \bibnamefont {Dzurak}},\ }\bibfield  {title} {\bibinfo {title} {Integrated silicon qubit platform with single-spin
  addressability, exchange control and single-shot singlet-triplet readout},\ }\href {https://doi.org/10.1038/s41467-018-06039-x} {\bibfield  {journal} {\bibinfo  {journal} {Nature Communications}\ }\textbf {\bibinfo {volume} {9}},\ \bibinfo {pages} {4370} (\bibinfo {year} {2018})}\BibitemShut {NoStop}%
\bibitem [{\citenamefont {Urdampilleta}\ \emph {et~al.}(2019)\citenamefont {Urdampilleta}, \citenamefont {Niegemann}, \citenamefont {Chanrion}, \citenamefont {Jadot}, \citenamefont {Spence}, \citenamefont {Mortemousque}, \citenamefont {Bäuerle}, \citenamefont {Hutin}, \citenamefont {Bertrand}, \citenamefont {Barraud}, \citenamefont {Maurand}, \citenamefont {Sanquer}, \citenamefont {Jehl}, \citenamefont {De~Franceschi}, \citenamefont {Vinet},\ and\ \citenamefont {Meunier}}]{Urdampilleta2019}%
  \BibitemOpen
  \bibfield  {author} {\bibinfo {author} {\bibfnamefont {M.}~\bibnamefont {Urdampilleta}}, \bibinfo {author} {\bibfnamefont {D.~J.}\ \bibnamefont {Niegemann}}, \bibinfo {author} {\bibfnamefont {E.}~\bibnamefont {Chanrion}}, \bibinfo {author} {\bibfnamefont {B.}~\bibnamefont {Jadot}}, \bibinfo {author} {\bibfnamefont {C.}~\bibnamefont {Spence}}, \bibinfo {author} {\bibfnamefont {P.-A.}\ \bibnamefont {Mortemousque}}, \bibinfo {author} {\bibfnamefont {C.}~\bibnamefont {Bäuerle}}, \bibinfo {author} {\bibfnamefont {L.}~\bibnamefont {Hutin}}, \bibinfo {author} {\bibfnamefont {B.}~\bibnamefont {Bertrand}}, \bibinfo {author} {\bibfnamefont {S.}~\bibnamefont {Barraud}}, \bibinfo {author} {\bibfnamefont {R.}~\bibnamefont {Maurand}}, \bibinfo {author} {\bibfnamefont {M.}~\bibnamefont {Sanquer}}, \bibinfo {author} {\bibfnamefont {X.}~\bibnamefont {Jehl}}, \bibinfo {author} {\bibfnamefont {S.}~\bibnamefont {De~Franceschi}}, \bibinfo {author} {\bibfnamefont {M.}~\bibnamefont {Vinet}},\ and\ \bibinfo {author}
  {\bibfnamefont {T.}~\bibnamefont {Meunier}},\ }\bibfield  {title} {\bibinfo {title} {Gate-based high fidelity spin readout in a cmos device},\ }\href {https://doi.org/10.1038/s41565-019-0443-9} {\bibfield  {journal} {\bibinfo  {journal} {Nature Nanotechnology}\ }\textbf {\bibinfo {volume} {14}},\ \bibinfo {pages} {737} (\bibinfo {year} {2019})}\BibitemShut {NoStop}%
\bibitem [{\citenamefont {Connors}\ \emph {et~al.}(2020)\citenamefont {Connors}, \citenamefont {Nelson},\ and\ \citenamefont {Nichol}}]{Connors2020}%
  \BibitemOpen
  \bibfield  {author} {\bibinfo {author} {\bibfnamefont {E.~J.}\ \bibnamefont {Connors}}, \bibinfo {author} {\bibfnamefont {J.}~\bibnamefont {Nelson}},\ and\ \bibinfo {author} {\bibfnamefont {J.~M.}\ \bibnamefont {Nichol}},\ }\bibfield  {title} {\bibinfo {title} {Rapid high-fidelity spin-state readout in $\mathrm{Si}$/$\mathrm{Si}$-$\mathrm{Ge}$ quantum dots via rf reflectometry},\ }\href {https://doi.org/10.1103/PhysRevApplied.13.024019} {\bibfield  {journal} {\bibinfo  {journal} {Phys. Rev. Appl.}\ }\textbf {\bibinfo {volume} {13}},\ \bibinfo {pages} {024019} (\bibinfo {year} {2020})}\BibitemShut {NoStop}%
\bibitem [{\citenamefont {Jang}\ \emph {et~al.}(2021)\citenamefont {Jang}, \citenamefont {Cho}, \citenamefont {Jang}, \citenamefont {Kim}, \citenamefont {Park}, \citenamefont {Kim}, \citenamefont {Kang}, \citenamefont {Jung}, \citenamefont {Umansky},\ and\ \citenamefont {Kim}}]{Jang2021}%
  \BibitemOpen
  \bibfield  {author} {\bibinfo {author} {\bibfnamefont {W.}~\bibnamefont {Jang}}, \bibinfo {author} {\bibfnamefont {M.-K.}\ \bibnamefont {Cho}}, \bibinfo {author} {\bibfnamefont {H.}~\bibnamefont {Jang}}, \bibinfo {author} {\bibfnamefont {J.}~\bibnamefont {Kim}}, \bibinfo {author} {\bibfnamefont {J.}~\bibnamefont {Park}}, \bibinfo {author} {\bibfnamefont {G.}~\bibnamefont {Kim}}, \bibinfo {author} {\bibfnamefont {B.}~\bibnamefont {Kang}}, \bibinfo {author} {\bibfnamefont {H.}~\bibnamefont {Jung}}, \bibinfo {author} {\bibfnamefont {V.}~\bibnamefont {Umansky}},\ and\ \bibinfo {author} {\bibfnamefont {D.}~\bibnamefont {Kim}},\ }\bibfield  {title} {\bibinfo {title} {Single-shot readout of a driven hybrid qubit in a gaas double quantum dot},\ }\href {https://doi.org/10.1021/acs.nanolett.1c00783} {\bibfield  {journal} {\bibinfo  {journal} {Nano Letters}\ }\textbf {\bibinfo {volume} {21}},\ \bibinfo {pages} {4999} (\bibinfo {year} {2021})}\BibitemShut {NoStop}%
\bibitem [{\citenamefont {Ma}\ \emph {et~al.}(2024)\citenamefont {Ma}, \citenamefont {Zhu}, \citenamefont {Kong}, \citenamefont {Sun}, \citenamefont {Ni}, \citenamefont {Zhou}, \citenamefont {Zhou}, \citenamefont {Luo}, \citenamefont {Cao}, \citenamefont {Wang}, \citenamefont {Li},\ and\ \citenamefont {Guo}}]{Ma2024}%
  \BibitemOpen
  \bibfield  {author} {\bibinfo {author} {\bibfnamefont {R.-L.}\ \bibnamefont {Ma}}, \bibinfo {author} {\bibfnamefont {S.-K.}\ \bibnamefont {Zhu}}, \bibinfo {author} {\bibfnamefont {Z.-Z.}\ \bibnamefont {Kong}}, \bibinfo {author} {\bibfnamefont {T.-P.}\ \bibnamefont {Sun}}, \bibinfo {author} {\bibfnamefont {M.}~\bibnamefont {Ni}}, \bibinfo {author} {\bibfnamefont {Y.-C.}\ \bibnamefont {Zhou}}, \bibinfo {author} {\bibfnamefont {Y.}~\bibnamefont {Zhou}}, \bibinfo {author} {\bibfnamefont {G.}~\bibnamefont {Luo}}, \bibinfo {author} {\bibfnamefont {G.}~\bibnamefont {Cao}}, \bibinfo {author} {\bibfnamefont {G.-L.}\ \bibnamefont {Wang}}, \bibinfo {author} {\bibfnamefont {H.-O.}\ \bibnamefont {Li}},\ and\ \bibinfo {author} {\bibfnamefont {G.-P.}\ \bibnamefont {Guo}},\ }\bibfield  {title} {\bibinfo {title} {Singlet-triplet-state readout in silicon metal-oxide-semiconductor double quantum dots},\ }\href {https://doi.org/10.1103/PhysRevApplied.21.034022} {\bibfield  {journal} {\bibinfo  {journal} {Phys. Rev. Appl.}\
  }\textbf {\bibinfo {volume} {21}},\ \bibinfo {pages} {034022} (\bibinfo {year} {2024})}\BibitemShut {NoStop}%
\bibitem [{\citenamefont {Kojima}\ \emph {et~al.}(2021)\citenamefont {Kojima}, \citenamefont {Nakajima}, \citenamefont {Noiri}, \citenamefont {Yoneda}, \citenamefont {Otsuka}, \citenamefont {Takeda}, \citenamefont {Li}, \citenamefont {Bartlett}, \citenamefont {Ludwig}, \citenamefont {Wieck},\ and\ \citenamefont {Tarucha}}]{Kojima2021}%
  \BibitemOpen
  \bibfield  {author} {\bibinfo {author} {\bibfnamefont {Y.}~\bibnamefont {Kojima}}, \bibinfo {author} {\bibfnamefont {T.}~\bibnamefont {Nakajima}}, \bibinfo {author} {\bibfnamefont {A.}~\bibnamefont {Noiri}}, \bibinfo {author} {\bibfnamefont {J.}~\bibnamefont {Yoneda}}, \bibinfo {author} {\bibfnamefont {T.}~\bibnamefont {Otsuka}}, \bibinfo {author} {\bibfnamefont {K.}~\bibnamefont {Takeda}}, \bibinfo {author} {\bibfnamefont {S.}~\bibnamefont {Li}}, \bibinfo {author} {\bibfnamefont {S.~D.}\ \bibnamefont {Bartlett}}, \bibinfo {author} {\bibfnamefont {A.}~\bibnamefont {Ludwig}}, \bibinfo {author} {\bibfnamefont {A.~D.}\ \bibnamefont {Wieck}},\ and\ \bibinfo {author} {\bibfnamefont {S.}~\bibnamefont {Tarucha}},\ }\bibfield  {title} {\bibinfo {title} {Probabilistic teleportation of a quantum dot spin qubit},\ }\href {https://doi.org/10.1038/s41534-021-00403-4} {\bibfield  {journal} {\bibinfo  {journal} {npj Quantum Information}\ }\textbf {\bibinfo {volume} {7}},\ \bibinfo {pages} {68} (\bibinfo {year}
  {2021})}\BibitemShut {NoStop}%
\bibitem [{\citenamefont {Geng}\ \emph {et~al.}(2025)\citenamefont {Geng}, \citenamefont {Kiczynski}, \citenamefont {Timofeev}, \citenamefont {Osika}, \citenamefont {Keith}, \citenamefont {Rowlands}, \citenamefont {Kranz}, \citenamefont {Rahman}, \citenamefont {Chung}, \citenamefont {Keizer}, \citenamefont {Gorman},\ and\ \citenamefont {Simmons}}]{Geng2025}%
  \BibitemOpen
  \bibfield  {author} {\bibinfo {author} {\bibfnamefont {H.}~\bibnamefont {Geng}}, \bibinfo {author} {\bibfnamefont {M.}~\bibnamefont {Kiczynski}}, \bibinfo {author} {\bibfnamefont {A.~V.}\ \bibnamefont {Timofeev}}, \bibinfo {author} {\bibfnamefont {E.~N.}\ \bibnamefont {Osika}}, \bibinfo {author} {\bibfnamefont {D.}~\bibnamefont {Keith}}, \bibinfo {author} {\bibfnamefont {J.}~\bibnamefont {Rowlands}}, \bibinfo {author} {\bibfnamefont {L.}~\bibnamefont {Kranz}}, \bibinfo {author} {\bibfnamefont {R.}~\bibnamefont {Rahman}}, \bibinfo {author} {\bibfnamefont {Y.}~\bibnamefont {Chung}}, \bibinfo {author} {\bibfnamefont {J.~G.}\ \bibnamefont {Keizer}}, \bibinfo {author} {\bibfnamefont {S.~K.}\ \bibnamefont {Gorman}},\ and\ \bibinfo {author} {\bibfnamefont {M.~Y.}\ \bibnamefont {Simmons}},\ }\bibfield  {title} {\bibinfo {title} {High-fidelity sub-microsecond single-shot electron spin readout above 3.5k},\ }\href {https://doi.org/10.1038/s41467-025-58279-3} {\bibfield  {journal} {\bibinfo  {journal} {Nature
  Communications}\ }\textbf {\bibinfo {volume} {16}},\ \bibinfo {pages} {3382} (\bibinfo {year} {2025})},\ \bibinfo {note} {published April 10, 2025}\BibitemShut {NoStop}%
\bibitem [{\citenamefont {Tsoukalas}\ \emph {et~al.}(2025)\citenamefont {Tsoukalas}, \citenamefont {von Lüpke}, \citenamefont {Orekhov}, \citenamefont {Hetényi}, \citenamefont {Seidler}, \citenamefont {Sommer}, \citenamefont {Kelly}, \citenamefont {Massai}, \citenamefont {Aldeghi}, \citenamefont {Pita-Vidal}, \citenamefont {Hendrickx}, \citenamefont {Bedell}, \citenamefont {Paredes}, \citenamefont {Schupp}, \citenamefont {Mergenthaler}, \citenamefont {Salis}, \citenamefont {Fuhrer},\ and\ \citenamefont {Harvey-Collard}}]{tsoukalas2025}%
  \BibitemOpen
  \bibfield  {author} {\bibinfo {author} {\bibfnamefont {K.}~\bibnamefont {Tsoukalas}}, \bibinfo {author} {\bibfnamefont {U.}~\bibnamefont {von Lüpke}}, \bibinfo {author} {\bibfnamefont {A.}~\bibnamefont {Orekhov}}, \bibinfo {author} {\bibfnamefont {B.}~\bibnamefont {Hetényi}}, \bibinfo {author} {\bibfnamefont {I.}~\bibnamefont {Seidler}}, \bibinfo {author} {\bibfnamefont {L.}~\bibnamefont {Sommer}}, \bibinfo {author} {\bibfnamefont {E.~G.}\ \bibnamefont {Kelly}}, \bibinfo {author} {\bibfnamefont {L.}~\bibnamefont {Massai}}, \bibinfo {author} {\bibfnamefont {M.}~\bibnamefont {Aldeghi}}, \bibinfo {author} {\bibfnamefont {M.}~\bibnamefont {Pita-Vidal}}, \bibinfo {author} {\bibfnamefont {N.~W.}\ \bibnamefont {Hendrickx}}, \bibinfo {author} {\bibfnamefont {S.~W.}\ \bibnamefont {Bedell}}, \bibinfo {author} {\bibfnamefont {S.}~\bibnamefont {Paredes}}, \bibinfo {author} {\bibfnamefont {F.~J.}\ \bibnamefont {Schupp}}, \bibinfo {author} {\bibfnamefont {M.}~\bibnamefont {Mergenthaler}}, \bibinfo {author}
  {\bibfnamefont {G.}~\bibnamefont {Salis}}, \bibinfo {author} {\bibfnamefont {A.}~\bibnamefont {Fuhrer}},\ and\ \bibinfo {author} {\bibfnamefont {P.}~\bibnamefont {Harvey-Collard}},\ }\href {https://arxiv.org/abs/2501.14627} {\bibinfo {title} {A dressed singlet-triplet qubit in germanium}} (\bibinfo {year} {2025}),\ \Eprint {https://arxiv.org/abs/2501.14627} {arXiv:2501.14627 [cond-mat.mes-hall]} \BibitemShut {NoStop}%
\bibitem [{\citenamefont {Corrigan}\ \emph {et~al.}(2023)\citenamefont {Corrigan}, \citenamefont {Dodson}, \citenamefont {Thorgrimsson}, \citenamefont {Neyens}, \citenamefont {Knapp}, \citenamefont {McJunkin}, \citenamefont {Coppersmith},\ and\ \citenamefont {Eriksson}}]{Corrigan2023}%
  \BibitemOpen
  \bibfield  {author} {\bibinfo {author} {\bibfnamefont {J.}~\bibnamefont {Corrigan}}, \bibinfo {author} {\bibfnamefont {J.~P.}\ \bibnamefont {Dodson}}, \bibinfo {author} {\bibfnamefont {B.}~\bibnamefont {Thorgrimsson}}, \bibinfo {author} {\bibfnamefont {S.~F.}\ \bibnamefont {Neyens}}, \bibinfo {author} {\bibfnamefont {T.~J.}\ \bibnamefont {Knapp}}, \bibinfo {author} {\bibfnamefont {T.}~\bibnamefont {McJunkin}}, \bibinfo {author} {\bibfnamefont {S.~N.}\ \bibnamefont {Coppersmith}},\ and\ \bibinfo {author} {\bibfnamefont {M.~A.}\ \bibnamefont {Eriksson}},\ }\bibfield  {title} {\bibinfo {title} {Latched readout for the quantum dot hybrid qubit},\ }\href {https://doi.org/10.1063/5.0130865} {\bibfield  {journal} {\bibinfo  {journal} {Applied Physics Letters}\ }\textbf {\bibinfo {volume} {122}},\ \bibinfo {pages} {074001} (\bibinfo {year} {2023})}\BibitemShut {NoStop}%
\bibitem [{\citenamefont {Hendrickx}\ \emph {et~al.}(2021)\citenamefont {Hendrickx}, \citenamefont {Lawrie}, \citenamefont {Russ}, \citenamefont {van Riggelen}, \citenamefont {de~Snoo}, \citenamefont {Schouten}, \citenamefont {Sammak}, \citenamefont {Scappucci},\ and\ \citenamefont {Veldhorst}}]{Hendrickx2021}%
  \BibitemOpen
  \bibfield  {author} {\bibinfo {author} {\bibfnamefont {N.~W.}\ \bibnamefont {Hendrickx}}, \bibinfo {author} {\bibfnamefont {W.~I.~L.}\ \bibnamefont {Lawrie}}, \bibinfo {author} {\bibfnamefont {M.}~\bibnamefont {Russ}}, \bibinfo {author} {\bibfnamefont {F.}~\bibnamefont {van Riggelen}}, \bibinfo {author} {\bibfnamefont {S.~L.}\ \bibnamefont {de~Snoo}}, \bibinfo {author} {\bibfnamefont {R.~N.}\ \bibnamefont {Schouten}}, \bibinfo {author} {\bibfnamefont {A.}~\bibnamefont {Sammak}}, \bibinfo {author} {\bibfnamefont {G.}~\bibnamefont {Scappucci}},\ and\ \bibinfo {author} {\bibfnamefont {M.}~\bibnamefont {Veldhorst}},\ }\bibfield  {title} {\bibinfo {title} {A four-qubit germanium quantum processor},\ }\href {https://doi.org/10.1038/s41586-021-03332-6} {\bibfield  {journal} {\bibinfo  {journal} {Nature}\ }\textbf {\bibinfo {volume} {591}},\ \bibinfo {pages} {580} (\bibinfo {year} {2021})},\ \bibinfo {note} {published March 1, 2021}\BibitemShut {NoStop}%
\bibitem [{\citenamefont {Marton}\ \emph {et~al.}(2023)\citenamefont {Marton}, \citenamefont {Sachrajda}, \citenamefont {Korkusinski}, \citenamefont {Bogan},\ and\ \citenamefont {Studenikin}}]{Marton2023}%
  \BibitemOpen
  \bibfield  {author} {\bibinfo {author} {\bibfnamefont {V.}~\bibnamefont {Marton}}, \bibinfo {author} {\bibfnamefont {A.}~\bibnamefont {Sachrajda}}, \bibinfo {author} {\bibfnamefont {M.}~\bibnamefont {Korkusinski}}, \bibinfo {author} {\bibfnamefont {A.}~\bibnamefont {Bogan}},\ and\ \bibinfo {author} {\bibfnamefont {S.}~\bibnamefont {Studenikin}},\ }\bibfield  {title} {\bibinfo {title} {Coherence characteristics of a gaas single heavy-hole spin qubit using a modified single-shot latching readout technique},\ }\bibfield  {journal} {\bibinfo  {journal} {Nanomaterials}\ }\textbf {\bibinfo {volume} {13}},\ \href {https://doi.org/10.3390/nano13050950} {10.3390/nano13050950} (\bibinfo {year} {2023})\BibitemShut {NoStop}%
\bibitem [{\citenamefont {Song}\ \emph {et~al.}(2024)\citenamefont {Song}, \citenamefont {Yun}, \citenamefont {Kim}, \citenamefont {Jang}, \citenamefont {Jang}, \citenamefont {Park}, \citenamefont {Cho}, \citenamefont {Sohn}, \citenamefont {Usami}, \citenamefont {Miyamoto}, \citenamefont {Itoh},\ and\ \citenamefont {Kim}}]{Song2024}%
  \BibitemOpen
  \bibfield  {author} {\bibinfo {author} {\bibfnamefont {Y.}~\bibnamefont {Song}}, \bibinfo {author} {\bibfnamefont {J.}~\bibnamefont {Yun}}, \bibinfo {author} {\bibfnamefont {J.}~\bibnamefont {Kim}}, \bibinfo {author} {\bibfnamefont {W.}~\bibnamefont {Jang}}, \bibinfo {author} {\bibfnamefont {H.}~\bibnamefont {Jang}}, \bibinfo {author} {\bibfnamefont {J.}~\bibnamefont {Park}}, \bibinfo {author} {\bibfnamefont {M.-K.}\ \bibnamefont {Cho}}, \bibinfo {author} {\bibfnamefont {H.}~\bibnamefont {Sohn}}, \bibinfo {author} {\bibfnamefont {N.}~\bibnamefont {Usami}}, \bibinfo {author} {\bibfnamefont {S.}~\bibnamefont {Miyamoto}}, \bibinfo {author} {\bibfnamefont {K.~M.}\ \bibnamefont {Itoh}},\ and\ \bibinfo {author} {\bibfnamefont {D.}~\bibnamefont {Kim}},\ }\bibfield  {title} {\bibinfo {title} {Coherence of a field gradient driven singlet-triplet qubit coupled to multielectron spin states in {${}^{28}$Si/SiGe}},\ }\href {https://doi.org/10.1038/s41534-024-00869-y} {\bibfield  {journal} {\bibinfo  {journal} {npj
  Quantum Information}\ }\textbf {\bibinfo {volume} {10}},\ \bibinfo {pages} {77} (\bibinfo {year} {2024})},\ \bibinfo {note} {published August 14, 2024}\BibitemShut {NoStop}%
\bibitem [{\citenamefont {Nakajima}\ \emph {et~al.}(2017)\citenamefont {Nakajima}, \citenamefont {Delbecq}, \citenamefont {Otsuka}, \citenamefont {Stano}, \citenamefont {Amaha}, \citenamefont {Yoneda}, \citenamefont {Noiri}, \citenamefont {Kawasaki}, \citenamefont {Takeda}, \citenamefont {Allison}, \citenamefont {Ludwig}, \citenamefont {Wieck}, \citenamefont {Loss},\ and\ \citenamefont {Tarucha}}]{Nakajima2017}%
  \BibitemOpen
  \bibfield  {author} {\bibinfo {author} {\bibfnamefont {T.}~\bibnamefont {Nakajima}}, \bibinfo {author} {\bibfnamefont {M.~R.}\ \bibnamefont {Delbecq}}, \bibinfo {author} {\bibfnamefont {T.}~\bibnamefont {Otsuka}}, \bibinfo {author} {\bibfnamefont {P.}~\bibnamefont {Stano}}, \bibinfo {author} {\bibfnamefont {S.}~\bibnamefont {Amaha}}, \bibinfo {author} {\bibfnamefont {J.}~\bibnamefont {Yoneda}}, \bibinfo {author} {\bibfnamefont {A.}~\bibnamefont {Noiri}}, \bibinfo {author} {\bibfnamefont {K.}~\bibnamefont {Kawasaki}}, \bibinfo {author} {\bibfnamefont {K.}~\bibnamefont {Takeda}}, \bibinfo {author} {\bibfnamefont {G.}~\bibnamefont {Allison}}, \bibinfo {author} {\bibfnamefont {A.}~\bibnamefont {Ludwig}}, \bibinfo {author} {\bibfnamefont {A.~D.}\ \bibnamefont {Wieck}}, \bibinfo {author} {\bibfnamefont {D.}~\bibnamefont {Loss}},\ and\ \bibinfo {author} {\bibfnamefont {S.}~\bibnamefont {Tarucha}},\ }\bibfield  {title} {\bibinfo {title} {Robust single-shot spin measurement with 99.5
  array},\ }\href {https://doi.org/10.1103/PhysRevLett.119.017701} {\bibfield  {journal} {\bibinfo  {journal} {Phys. Rev. Lett.}\ }\textbf {\bibinfo {volume} {119}},\ \bibinfo {pages} {017701} (\bibinfo {year} {2017})}\BibitemShut {NoStop}%
\bibitem [{\citenamefont {Kelly}\ \emph {et~al.}(2025)\citenamefont {Kelly}, \citenamefont {Massai}, \citenamefont {Hetényi}, \citenamefont {Pita-Vidal}, \citenamefont {Orekhov}, \citenamefont {Carlsson}, \citenamefont {Seidler}, \citenamefont {Tsoukalas}, \citenamefont {Sommer}, \citenamefont {Aldeghi}, \citenamefont {Bedell}, \citenamefont {Paredes}, \citenamefont {Schupp}, \citenamefont {Mergenthaler}, \citenamefont {Fuhrer}, \citenamefont {Salis},\ and\ \citenamefont {Harvey-Collard}}]{kelly2025}%
  \BibitemOpen
  \bibfield  {author} {\bibinfo {author} {\bibfnamefont {E.~G.}\ \bibnamefont {Kelly}}, \bibinfo {author} {\bibfnamefont {L.}~\bibnamefont {Massai}}, \bibinfo {author} {\bibfnamefont {B.}~\bibnamefont {Hetényi}}, \bibinfo {author} {\bibfnamefont {M.}~\bibnamefont {Pita-Vidal}}, \bibinfo {author} {\bibfnamefont {A.}~\bibnamefont {Orekhov}}, \bibinfo {author} {\bibfnamefont {C.}~\bibnamefont {Carlsson}}, \bibinfo {author} {\bibfnamefont {I.}~\bibnamefont {Seidler}}, \bibinfo {author} {\bibfnamefont {K.}~\bibnamefont {Tsoukalas}}, \bibinfo {author} {\bibfnamefont {L.}~\bibnamefont {Sommer}}, \bibinfo {author} {\bibfnamefont {M.}~\bibnamefont {Aldeghi}}, \bibinfo {author} {\bibfnamefont {S.~W.}\ \bibnamefont {Bedell}}, \bibinfo {author} {\bibfnamefont {S.}~\bibnamefont {Paredes}}, \bibinfo {author} {\bibfnamefont {F.~J.}\ \bibnamefont {Schupp}}, \bibinfo {author} {\bibfnamefont {M.}~\bibnamefont {Mergenthaler}}, \bibinfo {author} {\bibfnamefont {A.}~\bibnamefont {Fuhrer}}, \bibinfo {author} {\bibfnamefont
  {G.}~\bibnamefont {Salis}},\ and\ \bibinfo {author} {\bibfnamefont {P.}~\bibnamefont {Harvey-Collard}},\ }\href {https://arxiv.org/abs/2504.06898} {\bibinfo {title} {Identifying and mitigating errors in hole spin qubit readout}} (\bibinfo {year} {2025}),\ \Eprint {https://arxiv.org/abs/2504.06898} {arXiv:2504.06898 [cond-mat.mes-hall]} \BibitemShut {NoStop}%
\bibitem [{\citenamefont {Park}\ \emph {et~al.}(2024)\citenamefont {Park}, \citenamefont {Benson}, \citenamefont {Corrigan}, \citenamefont {Dodson}, \citenamefont {Coppersmith}, \citenamefont {Friesen},\ and\ \citenamefont {Eriksson}}]{park2024}%
  \BibitemOpen
  \bibfield  {author} {\bibinfo {author} {\bibfnamefont {S.}~\bibnamefont {Park}}, \bibinfo {author} {\bibfnamefont {J.}~\bibnamefont {Benson}}, \bibinfo {author} {\bibfnamefont {J.}~\bibnamefont {Corrigan}}, \bibinfo {author} {\bibfnamefont {J.~P.}\ \bibnamefont {Dodson}}, \bibinfo {author} {\bibfnamefont {S.~N.}\ \bibnamefont {Coppersmith}}, \bibinfo {author} {\bibfnamefont {M.}~\bibnamefont {Friesen}},\ and\ \bibinfo {author} {\bibfnamefont {M.~A.}\ \bibnamefont {Eriksson}},\ }\href {https://arxiv.org/abs/2408.15380} {\bibinfo {title} {Single-shot latched readout of a quantum dot qubit using barrier gate pulsing}} (\bibinfo {year} {2024}),\ \Eprint {https://arxiv.org/abs/2408.15380} {arXiv:2408.15380 [cond-mat.mes-hall]} \BibitemShut {NoStop}%
\bibitem [{\citenamefont {Martinis}(2015)}]{Martinis2015}%
  \BibitemOpen
  \bibfield  {author} {\bibinfo {author} {\bibfnamefont {J.~M.}\ \bibnamefont {Martinis}},\ }\bibfield  {title} {\bibinfo {title} {Qubit metrology for building a fault-tolerant quantum computer},\ }\href {https://doi.org/10.1038/npjqi.2015.5} {\bibfield  {journal} {\bibinfo  {journal} {npj Quantum Information}\ }\textbf {\bibinfo {volume} {1}},\ \bibinfo {pages} {15005} (\bibinfo {year} {2015})},\ \bibinfo {note} {published October 27, 2015}\BibitemShut {NoStop}%
\bibitem [{\citenamefont {Het\'enyi}\ and\ \citenamefont {Wootton}(2024)}]{Hetenyi2024}%
  \BibitemOpen
  \bibfield  {author} {\bibinfo {author} {\bibfnamefont {B.}~\bibnamefont {Het\'enyi}}\ and\ \bibinfo {author} {\bibfnamefont {J.~R.}\ \bibnamefont {Wootton}},\ }\bibfield  {title} {\bibinfo {title} {Tailoring quantum error correction to spin qubits},\ }\href {https://doi.org/10.1103/PhysRevA.109.032433} {\bibfield  {journal} {\bibinfo  {journal} {Phys. Rev. A}\ }\textbf {\bibinfo {volume} {109}},\ \bibinfo {pages} {032433} (\bibinfo {year} {2024})}\BibitemShut {NoStop}%
\bibitem [{\citenamefont {Shi}\ \emph {et~al.}(2012)\citenamefont {Shi}, \citenamefont {Simmons}, \citenamefont {Prance}, \citenamefont {Gamble}, \citenamefont {Koh}, \citenamefont {Shim}, \citenamefont {Hu}, \citenamefont {Savage}, \citenamefont {Lagally}, \citenamefont {Eriksson}, \citenamefont {Friesen},\ and\ \citenamefont {Coppersmith}}]{Shi2012}%
  \BibitemOpen
  \bibfield  {author} {\bibinfo {author} {\bibfnamefont {Z.}~\bibnamefont {Shi}}, \bibinfo {author} {\bibfnamefont {C.~B.}\ \bibnamefont {Simmons}}, \bibinfo {author} {\bibfnamefont {J.~R.}\ \bibnamefont {Prance}}, \bibinfo {author} {\bibfnamefont {J.~K.}\ \bibnamefont {Gamble}}, \bibinfo {author} {\bibfnamefont {T.~S.}\ \bibnamefont {Koh}}, \bibinfo {author} {\bibfnamefont {Y.-P.}\ \bibnamefont {Shim}}, \bibinfo {author} {\bibfnamefont {X.}~\bibnamefont {Hu}}, \bibinfo {author} {\bibfnamefont {D.~E.}\ \bibnamefont {Savage}}, \bibinfo {author} {\bibfnamefont {M.~G.}\ \bibnamefont {Lagally}}, \bibinfo {author} {\bibfnamefont {M.~A.}\ \bibnamefont {Eriksson}}, \bibinfo {author} {\bibfnamefont {M.}~\bibnamefont {Friesen}},\ and\ \bibinfo {author} {\bibfnamefont {S.~N.}\ \bibnamefont {Coppersmith}},\ }\bibfield  {title} {\bibinfo {title} {Fast hybrid silicon double-quantum-dot qubit},\ }\href {https://doi.org/10.1103/PhysRevLett.108.140503} {\bibfield  {journal} {\bibinfo  {journal} {Phys. Rev. Lett.}\ }\textbf
  {\bibinfo {volume} {108}},\ \bibinfo {pages} {140503} (\bibinfo {year} {2012})}\BibitemShut {NoStop}%
\bibitem [{\citenamefont {Neyens}\ \emph {et~al.}(2024)\citenamefont {Neyens}, \citenamefont {Zietz}, \citenamefont {Watson}, \citenamefont {Luthi}, \citenamefont {Nethwewala}, \citenamefont {George}, \citenamefont {Henry}, \citenamefont {Islam}, \citenamefont {Wagner}, \citenamefont {Borjans}, \citenamefont {Connors}, \citenamefont {Corrigan}, \citenamefont {Curry}, \citenamefont {Keith}, \citenamefont {Kotlyar}, \citenamefont {Lampert}, \citenamefont {Mądzik}, \citenamefont {Millard}, \citenamefont {Mohiyaddin}, \citenamefont {Pellerano}, \citenamefont {Pillarisetty}, \citenamefont {Ramsey}, \citenamefont {Savytskyy}, \citenamefont {Schaal}, \citenamefont {Zheng}, \citenamefont {Ziegler}, \citenamefont {Bishop}, \citenamefont {Bojarski}, \citenamefont {Roberts},\ and\ \citenamefont {Clarke}}]{Neyens2024}%
  \BibitemOpen
  \bibfield  {author} {\bibinfo {author} {\bibfnamefont {S.}~\bibnamefont {Neyens}}, \bibinfo {author} {\bibfnamefont {O.~K.}\ \bibnamefont {Zietz}}, \bibinfo {author} {\bibfnamefont {T.~F.}\ \bibnamefont {Watson}}, \bibinfo {author} {\bibfnamefont {F.}~\bibnamefont {Luthi}}, \bibinfo {author} {\bibfnamefont {A.}~\bibnamefont {Nethwewala}}, \bibinfo {author} {\bibfnamefont {H.~C.}\ \bibnamefont {George}}, \bibinfo {author} {\bibfnamefont {E.}~\bibnamefont {Henry}}, \bibinfo {author} {\bibfnamefont {M.}~\bibnamefont {Islam}}, \bibinfo {author} {\bibfnamefont {A.~J.}\ \bibnamefont {Wagner}}, \bibinfo {author} {\bibfnamefont {F.}~\bibnamefont {Borjans}}, \bibinfo {author} {\bibfnamefont {E.~J.}\ \bibnamefont {Connors}}, \bibinfo {author} {\bibfnamefont {J.}~\bibnamefont {Corrigan}}, \bibinfo {author} {\bibfnamefont {M.~J.}\ \bibnamefont {Curry}}, \bibinfo {author} {\bibfnamefont {D.}~\bibnamefont {Keith}}, \bibinfo {author} {\bibfnamefont {R.}~\bibnamefont {Kotlyar}}, \bibinfo {author} {\bibfnamefont {L.~F.}\
  \bibnamefont {Lampert}}, \bibinfo {author} {\bibfnamefont {M.~T.}\ \bibnamefont {Mądzik}}, \bibinfo {author} {\bibfnamefont {K.}~\bibnamefont {Millard}}, \bibinfo {author} {\bibfnamefont {F.~A.}\ \bibnamefont {Mohiyaddin}}, \bibinfo {author} {\bibfnamefont {S.}~\bibnamefont {Pellerano}}, \bibinfo {author} {\bibfnamefont {R.}~\bibnamefont {Pillarisetty}}, \bibinfo {author} {\bibfnamefont {M.}~\bibnamefont {Ramsey}}, \bibinfo {author} {\bibfnamefont {R.}~\bibnamefont {Savytskyy}}, \bibinfo {author} {\bibfnamefont {S.}~\bibnamefont {Schaal}}, \bibinfo {author} {\bibfnamefont {G.}~\bibnamefont {Zheng}}, \bibinfo {author} {\bibfnamefont {J.}~\bibnamefont {Ziegler}}, \bibinfo {author} {\bibfnamefont {N.~C.}\ \bibnamefont {Bishop}}, \bibinfo {author} {\bibfnamefont {S.}~\bibnamefont {Bojarski}}, \bibinfo {author} {\bibfnamefont {J.}~\bibnamefont {Roberts}},\ and\ \bibinfo {author} {\bibfnamefont {J.~S.}\ \bibnamefont {Clarke}},\ }\bibfield  {title} {\bibinfo {title} {Probing single electrons across 300-mm spin
  qubit wafers},\ }\href {https://doi.org/10.1038/s41586-024-07275-6} {\bibfield  {journal} {\bibinfo  {journal} {Nature}\ }\textbf {\bibinfo {volume} {629}},\ \bibinfo {pages} {80} (\bibinfo {year} {2024})}\BibitemShut {NoStop}%
\bibitem [{\citenamefont {George}\ \emph {et~al.}(2024)\citenamefont {George}, \citenamefont {Madzik}, \citenamefont {Henry}, \citenamefont {Wagner}, \citenamefont {Islam}, \citenamefont {Borjans}, \citenamefont {Connors}, \citenamefont {Corrigan}, \citenamefont {Curry}, \citenamefont {Harper}, \citenamefont {Keith}, \citenamefont {Lampert}, \citenamefont {Luthi}, \citenamefont {Mohiyaddin}, \citenamefont {Murcia}, \citenamefont {Nair}, \citenamefont {Nahm}, \citenamefont {Nethwewala}, \citenamefont {Neyens}, \citenamefont {Raharjo}, \citenamefont {Rogan}, \citenamefont {Savytskyy}, \citenamefont {Watson}, \citenamefont {Ziegler}, \citenamefont {Zietz}, \citenamefont {Pellerano}, \citenamefont {Pillarisetty}, \citenamefont {Bishop}, \citenamefont {Bojarski}, \citenamefont {Roberts},\ and\ \citenamefont {Clarke}}]{George2024}%
  \BibitemOpen
  \bibfield  {author} {\bibinfo {author} {\bibfnamefont {H.~C.}\ \bibnamefont {George}}, \bibinfo {author} {\bibfnamefont {M.~T.}\ \bibnamefont {Madzik}}, \bibinfo {author} {\bibfnamefont {E.~M.}\ \bibnamefont {Henry}}, \bibinfo {author} {\bibfnamefont {A.~J.}\ \bibnamefont {Wagner}}, \bibinfo {author} {\bibfnamefont {M.~M.}\ \bibnamefont {Islam}}, \bibinfo {author} {\bibfnamefont {F.}~\bibnamefont {Borjans}}, \bibinfo {author} {\bibfnamefont {E.~J.}\ \bibnamefont {Connors}}, \bibinfo {author} {\bibfnamefont {J.}~\bibnamefont {Corrigan}}, \bibinfo {author} {\bibfnamefont {M.}~\bibnamefont {Curry}}, \bibinfo {author} {\bibfnamefont {M.~K.}\ \bibnamefont {Harper}}, \bibinfo {author} {\bibfnamefont {D.}~\bibnamefont {Keith}}, \bibinfo {author} {\bibfnamefont {L.}~\bibnamefont {Lampert}}, \bibinfo {author} {\bibfnamefont {F.}~\bibnamefont {Luthi}}, \bibinfo {author} {\bibfnamefont {F.~A.}\ \bibnamefont {Mohiyaddin}}, \bibinfo {author} {\bibfnamefont {S.}~\bibnamefont {Murcia}}, \bibinfo {author} {\bibfnamefont
  {R.}~\bibnamefont {Nair}}, \bibinfo {author} {\bibfnamefont {R.}~\bibnamefont {Nahm}}, \bibinfo {author} {\bibfnamefont {A.}~\bibnamefont {Nethwewala}}, \bibinfo {author} {\bibfnamefont {B.}~\bibnamefont {Neyens}, \bibfnamefont {Samuel~Patra}}, \bibinfo {author} {\bibfnamefont {R.~D.}\ \bibnamefont {Raharjo}}, \bibinfo {author} {\bibfnamefont {C.}~\bibnamefont {Rogan}}, \bibinfo {author} {\bibfnamefont {R.}~\bibnamefont {Savytskyy}}, \bibinfo {author} {\bibfnamefont {T.~F.}\ \bibnamefont {Watson}}, \bibinfo {author} {\bibfnamefont {J.}~\bibnamefont {Ziegler}}, \bibinfo {author} {\bibfnamefont {O.~K.}\ \bibnamefont {Zietz}}, \bibinfo {author} {\bibfnamefont {S.}~\bibnamefont {Pellerano}}, \bibinfo {author} {\bibfnamefont {R.}~\bibnamefont {Pillarisetty}}, \bibinfo {author} {\bibfnamefont {N.~C.}\ \bibnamefont {Bishop}}, \bibinfo {author} {\bibfnamefont {S.~A.}\ \bibnamefont {Bojarski}}, \bibinfo {author} {\bibfnamefont {J.}~\bibnamefont {Roberts}},\ and\ \bibinfo {author} {\bibfnamefont {J.~S.}\ \bibnamefont
  {Clarke}},\ }\bibfield  {title} {\bibinfo {title} {12-spin-qubit arrays fabricated on a 300 mm semiconductor manufacturing line},\ }\href@noop {} {\bibfield  {journal} {\bibinfo  {journal} {Nano Letters}\ }\textbf {\bibinfo {volume} {25}},\ \bibinfo {pages} {793} (\bibinfo {year} {2024})}\BibitemShut {NoStop}%
\bibitem [{\citenamefont {Higginbotham}\ \emph {et~al.}(2014)\citenamefont {Higginbotham}, \citenamefont {Kuemmeth}, \citenamefont {Hanson}, \citenamefont {Gossard},\ and\ \citenamefont {Marcus}}]{Higginbotham2014}%
  \BibitemOpen
  \bibfield  {author} {\bibinfo {author} {\bibfnamefont {A.~P.}\ \bibnamefont {Higginbotham}}, \bibinfo {author} {\bibfnamefont {F.}~\bibnamefont {Kuemmeth}}, \bibinfo {author} {\bibfnamefont {M.~P.}\ \bibnamefont {Hanson}}, \bibinfo {author} {\bibfnamefont {A.~C.}\ \bibnamefont {Gossard}},\ and\ \bibinfo {author} {\bibfnamefont {C.~M.}\ \bibnamefont {Marcus}},\ }\bibfield  {title} {\bibinfo {title} {Coherent operations and screening in multielectron spin qubits},\ }\href {https://doi.org/10.1103/PhysRevLett.112.026801} {\bibfield  {journal} {\bibinfo  {journal} {Phys. Rev. Lett.}\ }\textbf {\bibinfo {volume} {112}},\ \bibinfo {pages} {026801} (\bibinfo {year} {2014})}\BibitemShut {NoStop}%
\bibitem [{\citenamefont {Simmons}\ \emph {et~al.}(2011)\citenamefont {Simmons}, \citenamefont {Prance}, \citenamefont {Van~Bael}, \citenamefont {Koh}, \citenamefont {Shi}, \citenamefont {Savage}, \citenamefont {Lagally}, \citenamefont {Joynt}, \citenamefont {Friesen}, \citenamefont {Coppersmith},\ and\ \citenamefont {Eriksson}}]{Simmons2011}%
  \BibitemOpen
  \bibfield  {author} {\bibinfo {author} {\bibfnamefont {C.~B.}\ \bibnamefont {Simmons}}, \bibinfo {author} {\bibfnamefont {J.~R.}\ \bibnamefont {Prance}}, \bibinfo {author} {\bibfnamefont {B.~J.}\ \bibnamefont {Van~Bael}}, \bibinfo {author} {\bibfnamefont {T.~S.}\ \bibnamefont {Koh}}, \bibinfo {author} {\bibfnamefont {Z.}~\bibnamefont {Shi}}, \bibinfo {author} {\bibfnamefont {D.~E.}\ \bibnamefont {Savage}}, \bibinfo {author} {\bibfnamefont {M.~G.}\ \bibnamefont {Lagally}}, \bibinfo {author} {\bibfnamefont {R.}~\bibnamefont {Joynt}}, \bibinfo {author} {\bibfnamefont {M.}~\bibnamefont {Friesen}}, \bibinfo {author} {\bibfnamefont {S.~N.}\ \bibnamefont {Coppersmith}},\ and\ \bibinfo {author} {\bibfnamefont {M.~A.}\ \bibnamefont {Eriksson}},\ }\bibfield  {title} {\bibinfo {title} {Tunable spin loading and ${T}_{1}$ of a silicon spin qubit measured by single-shot readout},\ }\href {https://doi.org/10.1103/PhysRevLett.106.156804} {\bibfield  {journal} {\bibinfo  {journal} {Phys. Rev. Lett.}\ }\textbf {\bibinfo
  {volume} {106}},\ \bibinfo {pages} {156804} (\bibinfo {year} {2011})}\BibitemShut {NoStop}%
\bibitem [{\citenamefont {Hutin}\ \emph {et~al.}(2019)\citenamefont {Hutin}, \citenamefont {Bertrand}, \citenamefont {Chanrion}, \citenamefont {Bohuslavskyi}, \citenamefont {Ansaloni}, \citenamefont {Yang}, \citenamefont {Michniewicz}, \citenamefont {Niegemann}, \citenamefont {Spence}, \citenamefont {Lundberg}, \citenamefont {Chatterjee}, \citenamefont {Crippa}, \citenamefont {Li}, \citenamefont {Maurand}, \citenamefont {Jehl}, \citenamefont {Sanquer}, \citenamefont {Gonzalez-Zalba}, \citenamefont {Kuemmeth}, \citenamefont {Niquet}, \citenamefont {De~Franceschi}, \citenamefont {Urdampilleta}, \citenamefont {Meunier},\ and\ \citenamefont {Vinet}}]{Hutin2019}%
  \BibitemOpen
  \bibfield  {author} {\bibinfo {author} {\bibfnamefont {L.}~\bibnamefont {Hutin}}, \bibinfo {author} {\bibfnamefont {B.}~\bibnamefont {Bertrand}}, \bibinfo {author} {\bibfnamefont {E.}~\bibnamefont {Chanrion}}, \bibinfo {author} {\bibfnamefont {H.}~\bibnamefont {Bohuslavskyi}}, \bibinfo {author} {\bibfnamefont {F.}~\bibnamefont {Ansaloni}}, \bibinfo {author} {\bibfnamefont {T.-Y.}\ \bibnamefont {Yang}}, \bibinfo {author} {\bibfnamefont {J.}~\bibnamefont {Michniewicz}}, \bibinfo {author} {\bibfnamefont {D.~J.}\ \bibnamefont {Niegemann}}, \bibinfo {author} {\bibfnamefont {C.}~\bibnamefont {Spence}}, \bibinfo {author} {\bibfnamefont {T.}~\bibnamefont {Lundberg}}, \bibinfo {author} {\bibfnamefont {A.}~\bibnamefont {Chatterjee}}, \bibinfo {author} {\bibfnamefont {A.}~\bibnamefont {Crippa}}, \bibinfo {author} {\bibfnamefont {J.}~\bibnamefont {Li}}, \bibinfo {author} {\bibfnamefont {R.}~\bibnamefont {Maurand}}, \bibinfo {author} {\bibfnamefont {X.}~\bibnamefont {Jehl}}, \bibinfo {author} {\bibfnamefont
  {M.}~\bibnamefont {Sanquer}}, \bibinfo {author} {\bibfnamefont {M.~F.}\ \bibnamefont {Gonzalez-Zalba}}, \bibinfo {author} {\bibfnamefont {F.}~\bibnamefont {Kuemmeth}}, \bibinfo {author} {\bibfnamefont {Y.-M.}\ \bibnamefont {Niquet}}, \bibinfo {author} {\bibfnamefont {S.}~\bibnamefont {De~Franceschi}}, \bibinfo {author} {\bibfnamefont {M.}~\bibnamefont {Urdampilleta}}, \bibinfo {author} {\bibfnamefont {T.}~\bibnamefont {Meunier}},\ and\ \bibinfo {author} {\bibfnamefont {M.}~\bibnamefont {Vinet}},\ }\bibfield  {title} {\bibinfo {title} {Gate reflectometry for probing charge and spin states in linear si mos split-gate arrays},\ }in\ \href {https://doi.org/10.1109/IEDM19573.2019.8993580} {\emph {\bibinfo {booktitle} {2019 IEEE International Electron Devices Meeting (IEDM)}}}\ (\bibinfo {year} {2019})\ pp.\ \bibinfo {pages} {37.7.1--37.7.4}\BibitemShut {NoStop}%
\bibitem [{\citenamefont {Bogan}\ \emph {et~al.}(2019)\citenamefont {Bogan}, \citenamefont {Studenikin}, \citenamefont {Korkusinski}, \citenamefont {Gaudreau}, \citenamefont {Zawadzki}, \citenamefont {Sachrajda}, \citenamefont {Tracy}, \citenamefont {Reno},\ and\ \citenamefont {Hargett}}]{Bogan2019}%
  \BibitemOpen
  \bibfield  {author} {\bibinfo {author} {\bibfnamefont {A.}~\bibnamefont {Bogan}}, \bibinfo {author} {\bibfnamefont {S.}~\bibnamefont {Studenikin}}, \bibinfo {author} {\bibfnamefont {M.}~\bibnamefont {Korkusinski}}, \bibinfo {author} {\bibfnamefont {L.}~\bibnamefont {Gaudreau}}, \bibinfo {author} {\bibfnamefont {P.}~\bibnamefont {Zawadzki}}, \bibinfo {author} {\bibfnamefont {A.}~\bibnamefont {Sachrajda}}, \bibinfo {author} {\bibfnamefont {L.}~\bibnamefont {Tracy}}, \bibinfo {author} {\bibfnamefont {J.}~\bibnamefont {Reno}},\ and\ \bibinfo {author} {\bibfnamefont {T.}~\bibnamefont {Hargett}},\ }\bibfield  {title} {\bibinfo {title} {Single hole spin relaxation probed by fast single-shot latched charge sensing},\ }\href {https://doi.org/10.1038/s42005-019-0113-0} {\bibfield  {journal} {\bibinfo  {journal} {Communications Physics}\ }\textbf {\bibinfo {volume} {2}},\ \bibinfo {pages} {17} (\bibinfo {year} {2019})}\BibitemShut {NoStop}%
\bibitem [{\citenamefont {Vahapoglu}\ \emph {et~al.}(2021)\citenamefont {Vahapoglu}, \citenamefont {Slack-Smith}, \citenamefont {Leon}, \citenamefont {Lim}, \citenamefont {Hudson}, \citenamefont {Day}, \citenamefont {Tanttu}, \citenamefont {Yang}, \citenamefont {Laucht}, \citenamefont {Dzurak},\ and\ \citenamefont {Pla}}]{Vahapoglu2021}%
  \BibitemOpen
  \bibfield  {author} {\bibinfo {author} {\bibfnamefont {E.}~\bibnamefont {Vahapoglu}}, \bibinfo {author} {\bibfnamefont {J.~P.}\ \bibnamefont {Slack-Smith}}, \bibinfo {author} {\bibfnamefont {R.~C.~C.}\ \bibnamefont {Leon}}, \bibinfo {author} {\bibfnamefont {W.~H.}\ \bibnamefont {Lim}}, \bibinfo {author} {\bibfnamefont {F.~E.}\ \bibnamefont {Hudson}}, \bibinfo {author} {\bibfnamefont {T.}~\bibnamefont {Day}}, \bibinfo {author} {\bibfnamefont {T.}~\bibnamefont {Tanttu}}, \bibinfo {author} {\bibfnamefont {C.~H.}\ \bibnamefont {Yang}}, \bibinfo {author} {\bibfnamefont {A.}~\bibnamefont {Laucht}}, \bibinfo {author} {\bibfnamefont {A.~S.}\ \bibnamefont {Dzurak}},\ and\ \bibinfo {author} {\bibfnamefont {J.~J.}\ \bibnamefont {Pla}},\ }\bibfield  {title} {\bibinfo {title} {Single-electron spin resonance in a nanoelectronic device using a global field},\ }\href {https://doi.org/10.1126/sciadv.abg9158} {\bibfield  {journal} {\bibinfo  {journal} {Science Advances}\ }\textbf {\bibinfo {volume} {7}},\ \bibinfo {pages}
  {eabg9158} (\bibinfo {year} {2021})}\BibitemShut {NoStop}%
\bibitem [{\citenamefont {Blumoff}\ \emph {et~al.}(2022)\citenamefont {Blumoff}, \citenamefont {Pan}, \citenamefont {Keating}, \citenamefont {Andrews}, \citenamefont {Barnes}, \citenamefont {Brecht}, \citenamefont {Croke}, \citenamefont {Euliss}, \citenamefont {Fast}, \citenamefont {Jackson}, \citenamefont {Jones}, \citenamefont {Kerckhoff}, \citenamefont {Lanza}, \citenamefont {Raach}, \citenamefont {Thomas}, \citenamefont {Velunta}, \citenamefont {Weinstein}, \citenamefont {Ladd}, \citenamefont {Eng}, \citenamefont {Borselli}, \citenamefont {Hunter},\ and\ \citenamefont {Rakher}}]{Blumoff2022}%
  \BibitemOpen
  \bibfield  {author} {\bibinfo {author} {\bibfnamefont {J.~Z.}\ \bibnamefont {Blumoff}}, \bibinfo {author} {\bibfnamefont {A.~S.}\ \bibnamefont {Pan}}, \bibinfo {author} {\bibfnamefont {T.~E.}\ \bibnamefont {Keating}}, \bibinfo {author} {\bibfnamefont {R.~W.}\ \bibnamefont {Andrews}}, \bibinfo {author} {\bibfnamefont {D.~W.}\ \bibnamefont {Barnes}}, \bibinfo {author} {\bibfnamefont {T.~L.}\ \bibnamefont {Brecht}}, \bibinfo {author} {\bibfnamefont {E.~T.}\ \bibnamefont {Croke}}, \bibinfo {author} {\bibfnamefont {L.~E.}\ \bibnamefont {Euliss}}, \bibinfo {author} {\bibfnamefont {J.~A.}\ \bibnamefont {Fast}}, \bibinfo {author} {\bibfnamefont {C.~A.}\ \bibnamefont {Jackson}}, \bibinfo {author} {\bibfnamefont {A.~M.}\ \bibnamefont {Jones}}, \bibinfo {author} {\bibfnamefont {J.}~\bibnamefont {Kerckhoff}}, \bibinfo {author} {\bibfnamefont {R.~K.}\ \bibnamefont {Lanza}}, \bibinfo {author} {\bibfnamefont {K.}~\bibnamefont {Raach}}, \bibinfo {author} {\bibfnamefont {B.~J.}\ \bibnamefont {Thomas}}, \bibinfo {author}
  {\bibfnamefont {R.}~\bibnamefont {Velunta}}, \bibinfo {author} {\bibfnamefont {A.~J.}\ \bibnamefont {Weinstein}}, \bibinfo {author} {\bibfnamefont {T.~D.}\ \bibnamefont {Ladd}}, \bibinfo {author} {\bibfnamefont {K.}~\bibnamefont {Eng}}, \bibinfo {author} {\bibfnamefont {M.~G.}\ \bibnamefont {Borselli}}, \bibinfo {author} {\bibfnamefont {A.~T.}\ \bibnamefont {Hunter}},\ and\ \bibinfo {author} {\bibfnamefont {M.~T.}\ \bibnamefont {Rakher}},\ }\bibfield  {title} {\bibinfo {title} {Fast and high-fidelity state preparation and measurement in triple-quantum-dot spin qubits},\ }\href {https://doi.org/10.1103/PRXQuantum.3.010352} {\bibfield  {journal} {\bibinfo  {journal} {PRX Quantum}\ }\textbf {\bibinfo {volume} {3}},\ \bibinfo {pages} {010352} (\bibinfo {year} {2022})}\BibitemShut {NoStop}%
\bibitem [{\citenamefont {Killick}\ \emph {et~al.}(2012)\citenamefont {Killick}, \citenamefont {Fearnhead},\ and\ \citenamefont {Eckley}}]{Killick2012}%
  \BibitemOpen
  \bibfield  {author} {\bibinfo {author} {\bibfnamefont {R.}~\bibnamefont {Killick}}, \bibinfo {author} {\bibfnamefont {P.}~\bibnamefont {Fearnhead}},\ and\ \bibinfo {author} {\bibfnamefont {I.~A.}\ \bibnamefont {Eckley}},\ }\bibfield  {title} {\bibinfo {title} {Optimal detection of changepoints with a linear computational cost},\ }\href {https://doi.org/10.1080/01621459.2012.737745} {\bibfield  {journal} {\bibinfo  {journal} {Journal of the American Statistical Association}\ }\textbf {\bibinfo {volume} {107}},\ \bibinfo {pages} {1590} (\bibinfo {year} {2012})}\BibitemShut {NoStop}%
\bibitem [{Mar()}]{Marciniec_zenodo}%
  \BibitemOpen
  \bibfield  {title} {\bibinfo {title} {Source data for the publication, ``{F}ast high- fidelity baseband reset of a latched state for quantum dot qubit readout," is available},\ }\href {https://doi.org/10.5281/zenodo.16616153} {10.5281/zenodo.16616153}\BibitemShut {NoStop}%
\end{thebibliography}%

\end{document}